\newif\ifpreprint

\preprintfalse 

\ifpreprint
\documentclass[journal=jctcce,manuscript=letter]{achemso}
\else
\documentclass[journal=jctcce,manuscript=letter,layout=twocolumn]{achemso}
\fi

\usepackage[T1]{fontenc} 

\usepackage{amsmath}
\usepackage{newtxtext,newtxmath}

\usepackage{graphicx}
\usepackage{dcolumn}
\usepackage{braket}
\usepackage{multirow}
\usepackage{threeparttable}
\usepackage{xspace}
\usepackage{verbatim}
\usepackage[version=4]{mhchem} 
\usepackage{comment}
\usepackage{color,soul}
\usepackage{siunitx}
\usepackage{physics}

\usepackage{mathtools}
\usepackage[dvipsnames]{xcolor}
\usepackage{xspace}
\usepackage{ifthen}

\usepackage{qcircuit}

\usepackage{graphicx,longtable,dcolumn,mhchem}
\usepackage{rotating,color}
\usepackage{lscape}
\usepackage{amsmath}
\usepackage{dsfont}
\usepackage{soul}
\usepackage{physics}
\newcolumntype{d}{D{.}{.}{-1}}

\newcommand{\pis}{\pi^\star}
\newcommand{\npi}{n\to\pis}
\newcommand{\ppi}{\pi\to\pis}

\usepackage[normalem]{ulem}

\usepackage[colorlinks = true,
            linkcolor = blue,
            urlcolor  = blue,
            citecolor = blue,
            anchorcolor = blue]{hyperref}
\definecolor{goodorange}{RGB}{225,125,0}
\definecolor{goodgreen}{RGB}{5,130,5}
\definecolor{goodred}{RGB}{220,50,25}
\definecolor{goodblue}{RGB}{30,144,255}

\newcommand{\note}[2]{
\ifthenelse{\equal{#1}{F}}{
\colorbox{goodorange}{\textcolor{white}{\footnotesize \fontfamily{phv}\selectfont #1}}
    \textcolor{goodorange}{{\footnotesize \fontfamily{phv}\selectfont #2}}\xspace
}{}
\ifthenelse{\equal{#1}{R}}{
\colorbox{goodred}{\textcolor{white}{\footnotesize \fontfamily{phv}\selectfont #1}}
    \textcolor{goodred}{{\footnotesize \fontfamily{phv}\selectfont #2}}\xspace
}{}
\ifthenelse{\equal{#1}{N}}{
\colorbox{goodgreen}{\textcolor{white}{\footnotesize \fontfamily{phv}\selectfont #1}}
    \textcolor{goodgreen}{{\footnotesize \fontfamily{phv}\selectfont #2}}\xspace
}{}
\ifthenelse{\equal{#1}{M}}{
\colorbox{goodblue}{\textcolor{white}{\footnotesize \fontfamily{phv}\selectfont #1}}
    \textcolor{goodblue}{{\footnotesize \fontfamily{phv}\selectfont #2}}\xspace
}{}
}

\usepackage{titlesec}

\usepackage[fontsize=11pt]{scrextend}
\titleformat{\section}
{\normalfont\sffamily\bfseries\color{Blue}}
{\thesection.}{0.25em}{\uppercase}

\titleformat{\subsection}
{\normalfont\sffamily\bfseries}
{\thesubsection}{0.25em}{}

\titleformat{\subsubsection}
{\normalfont\sffamily}
{\thesubsubsection}{0.25em}{}

\titleformat{\suppinfo}
{\normalfont\sffamily\bfseries}
{\thesubsection}{0.25em}{}

\titlespacing*{\section}{0pt}{0.5\baselineskip}{0.01\baselineskip}
\titlespacing*{\subsection}{0pt}{0.125\baselineskip}{0.01\baselineskip}
\titlespacing*{\subsubsection}{0pt}{0.125\baselineskip}{0.01\baselineskip}

\newcommand{\CEISAM}{Nantes Universit\'e, CNRS,  CEISAM UMR 6230, F-44000 Nantes, France}
\newcommand{\UOP}{Dipartimento di Chimica e Chimica Industriale, University of Pisa, Via Moruzzi 3, 56124 Pisa, Italy}

\newcommand{\LCPQ}{Laboratoire de Chimie et Physique Quantiques, Universit\'e de Toulouse, CNRS, {F-31062} Toulouse, France}
\newcommand{\IUF}{Institut Universitaire de France (IUF), F-75005 Paris, France}

\author{Pierre-Fran\c{c}ois Loos}
	\email{loos@irsamc.ups-tlse.fr}
	\affiliation[LCPQ, Toulouse]{\LCPQ}
\author{Martial Boggio-Pasqua}
	\affiliation[LCPQ, Toulouse]{\LCPQ}
\author{Aymeric Blondel}
	\affiliation[UN, Nantes]{\CEISAM}    
\author{Filippo Lipparini}
	\affiliation[UP, Pisa]{\UOP}
\author{Denis Jacquemin}
	\email{Denis.Jacquemin@univ-nantes.fr}
	\affiliation[UN, Nantes]{\CEISAM}    
	\alsoaffiliation[IUF, Paris]{\IUF}

\let\oldmaketitle\maketitle
\let\maketitle\relax
     \title{The QUEST Database of Highly-Accurate Excitation Energies}
\date{\today}

\begin{tocentry}
	\centering
	\includegraphics[width=.74\textwidth]{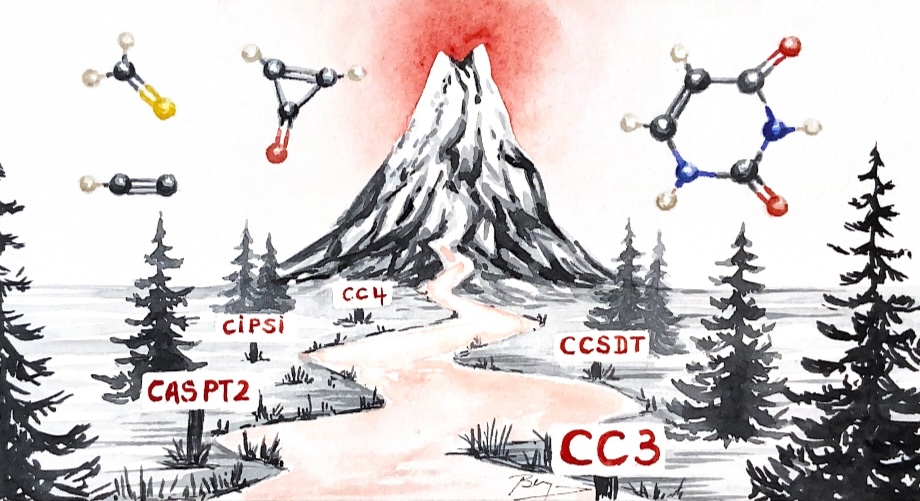}
\end{tocentry}

\begin{document}

\ifpreprint
\else
\twocolumn[
\begin{@twocolumnfalse}
\fi
\oldmaketitle

\begin{abstract}
We report theoretical best estimates of vertical transition energies (VTEs) for a large number of excited states and molecules: the \textsc{quest} database.  This database includes 1489  \emph{aug}-cc-pVTZ VTEs (731 singlets, 233 doublets, 461 triplets, and 64 quartets) for both valence and Rydberg transitions occurring in molecules containing from 1 to 16 non-hydrogen atoms.   \textsc{Quest} also includes a significant list of VTEs for states characterized by a partial or genuine double-excitation character, known to be particularly challenging for many computational methods. The vast majority of the reported values are deemed chemically-accurate, that is, are within $\pm0.05$ eV of the FCI/\emph{aug}-cc-pVTZ estimate. This allows for a balanced assessment of the performance of popular excited-state methodologies. We report the results of such benchmarks for various single- and multi-reference wavefunction approaches, and provide extensive supporting information allowing testing of other models. All corresponding data associated with the \textsc{quest} database, along with analysis tools, can be found in the associated \textsc{GitHub} repository at the following URL: \url{https://github.com/pfloos/QUESTDB}.
\end{abstract}

\ifpreprint
\else
\end{@twocolumnfalse}
]
\fi

\ifpreprint
\else
\small
\fi

\noindent

\section{Introduction}
\label{sec:Intro}

Ongoing advances in electronic structure theories are continually expanding the scope of problems that theoretical chemists can tackle. As more efficient and accurate methods become available, questions of chemical relevance that were computationally inaccessible only a few years ago are now within reach. However, assessing the reliability of new approaches typically requires extensive benchmarking, since the performance of a given method can vary significantly, not only across different molecular families (e.g., organic \textit{vs} inorganic systems) but also depending on the property of interest (e.g., geometries \textit{vs} spectroscopic signatures). This variability is exemplified by the well-known difficulty of simultaneously achieving accurate energies and electron densities within density-functional theory (DFT). \cite{Med17,Kep17,Ham17,Kor17} Reliable reference values are essential for benchmarking, and numerous datasets, based on either experimental measurements or high-level theoretical calculations, have been developed and are now widely used. \cite{Pop89,Cur91,Cur97,Cur98,Fur02,Die04,Die04b,Taj04,Bom06,Zha06i,Jur06,Zha06i,Cur07b,Sch08,Sil08,Roh08,Har08,Pea08,Zha08b,Ver09,Car10b,Sil10,Sil10b,Sil10c,Goe10,Goe10a,Goe11,Goe11b,Goe11c,Sen11,Sen11b,Sen11c,Rez11,Jac12c,Jac12d,Ise12,Lea12,Gui13,Win13,Fil13,Har14,Kan14,Mew15,Van15,Jac15a,Jac15b,Li16a,Li16,Mot17,Goe17,Mar17c,Sch17,Bud17,Cas19,Cas20,Koz20,Wil20,Stu20,Fra21,Van22,Son25} 

In the context of excited-state (ES) calculations, relying on experimental reference values presents at least two major limitations. First, many ESs are either dark or nearly dark in one-photon spectroscopies, meaning they do not appear in conventional absorption experiments. Detecting and characterizing such states often requires more advanced techniques, which have only been successfully applied to a quite limited number of molecules and transitions. As a result, available experimental data are inherently biased toward low-lying bright excited states. Second, the most readily computed feature in theoretical ES studies, the vertical transition energy (VTE), typically lacks a direct experimental counterpart. Measured quantities such as the absorption maximum ($\lambda_\text{max}$) or the 0-0 energy include vibrational effects that are absent in VTE calculations. Consequently, it is not surprising that theoreticians have mainly focused on establishing theoretical reference for VTEs.

In the early 1990s, Roos and collaborators pioneered the multireference (MR) complete active space second-order perturbation theory (CASPT2), \cite{And90,And92} applying it to compute accurate VTEs for a broad range of small molecules. \cite{Ser93,Ser93b,Ser95,Lor95b,Ser96b} Around the same time, Bartlett and co-workers developed a family of single-reference (SR) coupled-cluster (CC) methods and employed them to investigate excited states, often targeting the same systems studied by Serrano-Andr\'es and Roos.\cite{Gwa95,Wat96b,Del97b} While both approaches achieved notable success, significant discrepancies arose in their predictions for certain ESs. At the time, however, the absence of indisputable theoretical reference values made it difficult to determine the source of these differences.

\begin{table*}[htp]
\caption{\footnotesize Previous publications of the \textsc{quest} database (in chronological order) summarizing the nature of the investigated molecules and ESs together with the main electronic structure methods employed to establish the TBEs.$^a$ 
}
\label{Table-1}
\vspace{-0.3 cm}
\footnotesize
\begin{tabular}{lllllll}  
\hline
Year 	&Ref.				& Molecules			& States/Properties		& \# of ESs	& Reference method 	& Benchmarked methods\\
\hline
2018	&\citenum{Loo18a}		& 18 small 				& --					& 110		&SCI						&SR-$\Psi$						\\
2019	&\citenum{Loo19c}		& 14 small and medium		& Double ESs			& 20			&SCI						&SR-$\Psi$ \& MR-$\Psi$			\\
2020	&\citenum{Loo20a}		& 27 medium				& --					& 238		&CCSDTQ \& CCSDT 		&SR-$\Psi$						\\
2020	&\citenum{Loo20b}		& 45 small and medium		& --					& 328		&--						&ADC(2) \& ADC(3)				\\
2020	&\citenum{Loo20d}		& 14 small exotic			& --					& 30			&SCI						&SR-$\Psi$						\\
2020	&\citenum{Loo20d}		& 24 small radicals			& Doublet ESs			& 51			&SCI						&SR-$\Psi$						\\
2021	&\citenum{Chr21}$^b$	& 13 small					& $f$ and $\mu$		&30			&CCSDTQP				&--								\\
2021	&\citenum{Sar21}$^b$	& 13 small					& $f$ and $\mu$		&30			&--						&SR-$\Psi$ \& TD-DFT			\\	
2021	&\citenum{Ver21}$^c$	& 13 medium and large		&--					&119			&CCSDT					&SR-$\Psi$						\\
2021	&\citenum{Loo21a}		& 17 medium 				& Charge-transfer ESs	&30			& CCSDT					&SR-$\Psi$, TD-DFT \& BSE/$GW$	\\
2021	&\citenum{Loo21}		& 10 bicyclic systems		& --					&91			&CCSDT \& CC3			&SR-$\Psi$						\\
2021	&\citenum{Loo21b}		& 10 small					& --					&28			&--						&CC4							\\
2022	&\citenum{Loo22}$^d$	& 43 small and medium		& --					&84 (200)	&CC4						&None							\\
2022	&\citenum{Sar22}		& 35 small and medium		& --					& 284		& --						&MR-$\Psi$						\\
2022	&\citenum{Bog22}		& 35 small and medium		&--					&220			&--						&CASPT3							\\
2023	&\citenum{Loo23a}		& 10 triangulenes			& Singlet-triplet gaps		&20			&CCSDT \& CC3			&SR-$\Psi$ \& TD-DFT			\\	
2023	&\citenum{Jac23a}		& 11 metallic diatomics		& --					&67			& SCI, CCSDTQ \& CASPT2	&SR-$\Psi$ \& MR-$\Psi$			\\
2024	&\citenum{Loo24}		& 12 substituted benzenes	&--					&108			&CCSDT					&SR-$\Psi$						\\
2024	&\citenum{Kos24}		& 26 small and medium		& Double ESs			& 47			&SCI, CC4 \& CASPT3		&SR-$\Psi$ \& MR-$\Psi$			\\
2024	&\citenum{Kny24}		& 13 large chromophores		& --					& 120		&CC3					&SR-$\Psi$, TD-DFT \& BSE/$GW$	\\
2024	&\citenum{Nai24}$^b$	& 25 small and medium		& $\sigma^\text{TPA}$	& 92			&CC3					&SR-$\Psi$ \& TD-DFT			\\
2025	&\citenum{Sir25}$^b$	& 21 small and medium		& $f^\text{ESA}$		& 71			&CC3					&SR-$\Psi$ \& TD-DFT			\\
\hline
\end{tabular}
\vspace{-0.3 cm}
\begin{flushleft}
$^a$SR-$\Psi$ and MR-$\Psi$ stand for single- and multi-reference wave function methods, respectively; $f$, $\mu$, $\sigma^\text{TPA}$, and  $f^\text{ESA}$ are the oscillator strength, dipole moment, two-photon absorption cross-section, and ES absorption oscillator strength, respectively. $^b$ These studies are focused on the computation of properties rather than energies. $^c$ Review of our earlier works. $^d$ Reinvestigation of approximately 200 previously defined TBEs with CC4 leading to 84 improved reference values.
\end{flushleft}
\end{table*}

Starting in 2008, the Thiel group established a widely used set of theoretical best estimates (TBEs) for VTEs. To construct this benchmark, they examined 28 CNOH-containing $\pi$-conjugated molecules, including DNA bases and naphthalene, producing 104 singlet and 63 triplet reference values for valence ESs.\cite{Sch08} Notably, based on second-order M{\o}ller-Plesset geometries, Thiel and collaborators mainly employed CASPT2 and third-order coupled-cluster (CC3) \cite{Chr95b,Koc95,Koc97} to derive these TBEs. Two versions of the dataset were released, mainly differing by the atomic basis sets used (TZVP \textit{vs} aug-cc-pVTZ) and, to some extent, by the reference methods applied. \cite{Sch08,Sil10b,Sil10c} It is noteworthy that a significant fraction of the TBEs selected by Thiel's group originated from earlier studies by other groups, which employed different geometries, electronic structure methods, and basis sets than those used in Refs.~\citenum{Sch08,Sil10b,Sil10c} to define their values. For reasons that remain unclear to us, this aspect, although clearly stated in Thiel's work, appears to have been overlooked in many subsequent studies relying on Thiel's TBEs. Indeed, Thiel's benchmark set quickly gained traction in the electronic structure community, accumulating over 1,000 citations according to Google Scholar (including around 50 in 2024), underscoring its broad impact. This popularity was anticipated by the authors, who stated in their seminal publication: \textit{``We expect this benchmark set to be useful for validation and development purposes, and anticipate future improvements and extensions of this set through further high-level calculations.''} \cite{Sch08}

A decade later, building on more advanced theoretical methods, software, and computational resources than were available to our predecessors, we set out to fulfill the second part of Thiel's prediction: establishing a new set of reference VTEs using high-level approaches. Our initial objective was to produce indisputable VTEs with a target error margin of 0.05 eV (equivalent to approximately 1 kcal/mol), commonly referred to as chemical accuracy. To achieve this, we employed a combination of high-level CC methods, \cite{Nog87,Scu88,Chr95b,Koc95,Koc97,Kuc01,Kow01,Kow01b,Hir04,Kal03,Kal04,Kal05} selected configuration interaction (SCI), \cite{Hur73,Gin13,Gin15,Caf16,Gar17,Gar18,Gar19,Loo20a,Loo20b,Dam21,Dam22} and a variety of MR approaches. \cite{And90,And92,Wer96,Ang01b,Ang02} In principle, each of these techniques should converge to the same estimate, provided that sufficiently high excitation levels, extensive CI expansions, and large active spaces are used, respectively. This combined human and computational effort, comparable to a team mountaineering expedition, ultimately exceeded our original expectations from 2018 and culminated in the creation of a new benchmark dataset: the \textsc{quest} database.

As an initial overview of the \textsc{quest} project, Table \ref{Table-1} lists our previous contributions,\cite{Loo18a,Loo19c,Loo20a,Loo20b,Loo20d,Chr21,Sar21,Ver21,Loo21a,Loo21,Loo21b,Loo22,Sar22,Bog22,Loo23a,Jac23a,Loo24,Kos24,Kny24,Nai24,Sir25} that have led to the current version of the database. While most of these works involve both the definition of new TBEs and systematic benchmarking, a few of them focus more narrowly on one of these aspects. To the best of our knowledge, only one substantial external extension of our TBEs has been proposed to date: the 2025 study by Song and co-workers, which specifically targets open-shell molecules. \cite{Son25} In contrast, a growing number of research groups have already employed all or part of the \textsc{quest} dataset to 
assess the accuracy of various theoretical methods. \cite{Gin19,Cas19,Oti20,Hai20b,Hai21,Gou21,Gro21,Mes21,Cas21b,Mes21b,Lia22,Mes22,Gou22,Kin22,Van22,Fur23,Hen23,Kin23,Gro23,Ent23,Ris23,Zie23,Dat24,Cas24,Cos24,Pfa24,Sza24,Sta24,Gou24,Bin25,For25,Mil25,Tra25,Yu25,Son25,Val25}

In this contribution, our objective is to present all the VTEs we have accumulated to date in a format that is as consistent and uniform as possible, together with benchmark results for various wavefunction-based methods using the \textsc{quest} TBEs. Importantly, many of the earlier TBEs have been re-evaluated using more advanced methodologies than those employed in our initial studies. In addition, the present work includes a significantly broader range of molecular structures and ESs. As such, it would be misleading to regard this contribution as a mere compilation of the studies listed in Table \ref{Table-1}. In other words, the data provided in the Supporting Information (SI) should be considered the most up-to-date, and we hope rather definitive, representation of our current reference values.

\section{Overview of the \textsc{quest} database}

\subsection{How are the TBEs defined?}

As is common practice in the field, \cite{Kal04,Bal06,Kam06b,Pea12,Wat12,Fel14,Gui18b,Fra19,Cas19} we typically apply additive basis set corrections to establish our TBEs, i.e., 
\begin{equation}
	\label{eq1}
	\Delta {E}_{\text{Large }}^{\text{TBE}}  =  \Delta E_{\text{Small }}^{\text{High}}  +  \qty[ \Delta E_{\text{Large }}^{\text{Low}} 
	- \Delta E_{\text{Small }}^{\text{Low}}  ], 
\end{equation}
where the subscripts and superscripts indicate the level of theory and basis set used, respectively. As detailed below, the error introduced by this approximation is often well-controlled when the ``low-level'' method already provides a reasonably accurate description of the ES under consideration. 

In \textsc{quest}, the large basis sets in Eq.~\eqref{eq1} are Dunning's \emph{aug}-cc-pVTZ (AVTZ) \cite{Dun89,Ken92,Won93,Pra11} or \emph{aug}-cc-pVQZ (AVQZ).   \cite{Dun89,Ken92,Won93,Pra11} \textsc{Quest}'s TBEs are systematically provided in the former, and, when possible, in the latter as well. As small basis sets we rely on the double-$\zeta$  \emph{aug}-cc-pVTZ (AVDZ)  basis \cite{Dun89,Ken92,Won93,Pra11} and 6-31+G(d). \cite{Dit71,Heh72,Har73,Cla83} As can be seen, only basis sets containing both diffuse and polarization functions have been selected, since both families of functions are often needed for describing ESs.

In our case, the low-level method in Eq.~\eqref{eq1} is, in the vast majority of cases, at least CC3, \cite{Chr95b,Koc95,Koc97} though we occasionally had to step down to CCSDR(3) \cite{Chr96b} or even CCSD \cite{Pur82,Scu87,Koc90b,Sta93,Sta93b} for the largest compounds. For the high-level method, we relied on the most accurate yet technically feasible approach, which naturally depends on the molecular size.

Let us elaborate on the key case of singlet ESs in closed-shell organic molecules. For tiny systems (typically with 1--2 non-H atoms), we used highly converged SCI calculations based on the Configuration Interaction using a Perturbative Selection made Iteratively (CIPSI) algorithm, \cite{Hur73,Gin13,Gin15,Gar17b,Gar18,Gar19} provided the extrapolation error was below 0.010 eV. In such cases, SCI served as our reference. Full configuration interaction (FCI)-quality results could also be obtained using CCSDTQP, \cite{Kuc91,Hir04,Kal03,Kal04,Kal05} which we applied when possible.

For molecules containing 3--4 non-H atoms, CCSDTQ calculations in a double-$\zeta$ basis were generally achievable and used as the high-level reference. For slightly larger systems (5--7 non-H atoms), we typically employed CC4, \cite{Kal04,Kal05,Loo21b,Loo22} which includes iterative quadruples while scaling more favorably than CCSDTQ. In the case of even larger molecules (8--15 non-H atoms), the high-level method was limited to CCSDT \cite{Nog87,Scu88,Kuc01,Kow01,Kow01b} or CC3. \cite{Chr95b,Koc95,Koc97}

For excited states with significant double-excitation character in large molecules, reference values were established using MR approaches, typically NEVPT2 \cite{Ang01b,Ang02} or CASPT3. \cite{Wer96} These guidelines are general, and the specific method used to determine the TBE for each ES is provided in the SI.

For open-shell species, which are both computationally and chemically more challenging, we established the TBEs using the same strategy and basis sets as for closed-shell systems, but restricted our attention to compact molecules containing 1--4 non-H atoms. This allowed us to perform reference calculations using CCSDTQP or SCI for the smallest cases (1--2 non-H atoms), and CCSDTQ for slightly larger species (3--4 non-H atoms). These calculations were carried out using the restricted open-shell (RO) formalism. Note, however, that the unrestricted (U) formalism was employed in Ref.~\citenum{Loo20b}. In any case, both formalisms converge to the same VTEs at sufficiently high coupled-cluster levels, and in practice, differences are trifling when using CCSDTQ or CCSDTQP.

\subsection{State identification and characterizations}

\begin{table*}[htp]
\caption{\footnotesize TBEs of VTEs included in the \textsc{quest} subsets and the full database presented by ES types. S, T, D, and Q refer to singlet, triplet, doublet, and quartet states, respectively. GD, PD, CT, and FL denote genuine double, partial double, charge transfer, and fluorescence, respectively. }
\label{Table-2}
\vspace{-0.3 cm}
\footnotesize
\begin{tabular}{lcccccccccccccccccc}
\hline
			&			&		&		& 				\multicolumn{14}{c}{Number of excited states}\\
			\cline{6-19}
			& \multicolumn{2}{c}{Compounds}	&	\multicolumn{2}{c}{Transitions}			& \multicolumn{4}{c}{Size in non-H atoms$^a$} 	& \multicolumn{4}{c}{Spin}				&  \multicolumn{2}{c}{Nature}		&   \multicolumn{4}{c}{Particularity}\\
			& Total	 &\multicolumn{1}{c}{Nature}					&  Total & Safe		&1--2	& 3--5	& 6--9	& 10--16		& S	& T		& D	& Q	&  Val.			&  Ryd.	& GD		& PD		& CT		& FL\\
\hline
\textsc{main}	& 117 &Organic \& Inorg.		&  927 & 837		&129		&318		&338		&142			&582		&345		&		&		&659$^c$		&259	& 28		& 21		&28		&10\\
\textsc{rad}	& 33 & Open-shell			& 281 & 225		&201		&80		&		&			&		&		&217		&64		&166			&82		&11		&22		&		& \\
\textsc{chrom}	& 18 &Org. Chromophore		& 158 & $\sim$135$^b$&		&		&		&158			&86		&72		&		&		& 149$^d$	&9		& 		&7		&		&\\
\textsc{bio}	& 5 	&Nucleobases			& 56 & $\sim$51$^b$&		&		& 33		&23			&35		&21		&		&		&40$^e$		&16		&		&		&		&\\
\textsc{tm}		& 11 &Transition metal diatomics& 67 & 46		& 67		&		&		&			&28		&23		&16		&		&			&		& 4		&		&		&\\
\hline
\textsc{quest}	&184 &					& 1489 & $\sim$1294	&397		&398		&371		&323			&731		&461		&233		&64		&1014		&366		&43		&50		&28		&10	\\
\hline
\end{tabular}
\vspace{-0.3 cm}
\begin{flushleft}
$^a$ These four categories are denoted, tiny, small, medium, and large in the following.
$^b$ Since we do not provide safe/unsafe flags for these series (see text), these values correspond to the number of ESs having $\%T_1 > 85\%$.
$^c$ 386 $\ppi$ transitions and 207 $\npi$ transitions.
$^d$ 40 $\ppi$ transitions and 103 $\npi$ transitions.
$^e$ 25 $\ppi$ transitions and 16 $\npi$ transitions.
\end{flushleft}
\end{table*}

Clearly identifying each ES is crucial in benchmarking studies. For this reason, the database includes detailed information to unambiguously identify each state, thereby helping to avoid inappropriate comparisons, such as mixing fundamentally different types of transitions, during the benchmarking process.

Correspondences between ESs obtained with different quantum chemical methods and basis sets can generally be established without difficulty. To do so, we primarily relied on spatial and spin symmetries, as well as oscillator strengths (when non-zero) and dominant orbital transitions. It is worth noting that different software packages may use varying labels for certain spatial symmetries. For example, $B_{1u}$, $B_{2u}$, and $B_{3u}$ are interchangeable in $D_{2h}$ compounds. In such cases, orbital numbering provides a reliable means of identifying equivalent ESs.

For radicals, ES identification is more challenging due to several factors, \cite{Smi05b} including a generally larger contribution from double excitations and spin contamination when using SR wavefunction methods. Nonetheless, the same identification strategy was applied, with explicit consideration of both spin-up and spin-down orbital contributions.

The Rydberg or valence character of each ES was assessed by examining the dominant virtual orbitals and the spatial extent of the electronic cloud, as indicated by the expectation value $\expval*{r^2}$. The degree of double excitation was quantified, whenever possible, using the $\%T_1$ metric, which reflects the proportion of single excitations involved in the transition computed at the CC3 level. States with a dominant doubly-excited character, or ``genuine doubles'' (GD), are characterized by $\%T_1 < 50\%$, with most cases close to 0\%. States with a significant but not dominant doubly-excited character, or ``partial doubles'' (PD), typically corresponding to $\%T_1$ values in the range 60--80\%. A very small number of transitions were computed at the $S_1$ minimum geometry (i.e., relevant for fluorescence) and are indicated by the  ``FL'' tag. Several ESs show a weak or strong charge-transfer (CT) character and are labeled as wCT or sCT, respectively. There are various methods to quantify charge transfer,\cite{Pla20} and we employed three different metrics, described in more detail in a dedicated contribution. \cite{Loo21a} To assist users in identifying ESs using their preferred methodology, the \textsc{quest} database provides $\%T_1$, oscillator strength $f$, $\expval*{r^2}$, and dominant orbital contributions for all ESs, computed with reasonably high levels of theory (see the SI for details). Additionally, $\expval*{\Hat{S}^2}$ values are provided for doublet and quartet states, along with the dominant spin-up and spin-down orbital combinations computed at the CCSD level. In the following, S, T, D, and Q refer to singlet, triplet, doublet, and quartet states, respectively.

Of course, for the most challenging cases, these indicators can significantly change when altering the basis set and/or the electronic structure level. A few ESs are difficult to describe unambiguously or to trace consistently across all tested methods. Specifically, when ESs become nearly degenerate at a given (low) level of theory but not at the reference level, state-mixing can occur, making assignments dependent on the selected criterion. However, the assignments remain clear and straightforward in the vast majority of cases. In \textsc{quest}, the challenging cases are few and typically arise for ESs that are close in energy. These cases have a negligible statistical impact on the trends discussed below.

\subsection{Are \textsc{quest}'s TBEs chemically accurate?}

As stated in Section \ref{sec:Intro}, one of our original goals was to reach \emph{chemical accuracy}, that is, to provide TBEs with error bars smaller than $\pm 0.050$ eV within a given atomic basis set and at a fixed geometry. When both CIPSI and high-level CC calculations are well converged and yield negligible differences, meeting this objective is straightforward. In fact, throughout the \textsc{quest} collaboration, each partner was able to identify occasional errors in the results of the other team(s), including incorrect assignments, overlooked transitions, or simple typos. 

For larger systems, determining whether a TBE is ``safe'' (i.e., chemically accurate) becomes less straightforward, as the safety net provided by CIPSI, CCSDTQP, or MR calculations is no longer tight enough to rigorously assess the accuracy of, e.g., CCSDT predictions. In such cases, the safe/unsafe flags were assigned on a case-by-case basis, considering the $\%T_1$ diagnostic, the convergence behavior of the CC hierarchy, and the magnitude of basis set effects. For ESs with $\%T_1 \gtrsim 85\%$, small CC3-CCSDT differences, and conventional basis set effects, it appears reasonable to rate the TBEs as safe (\emph{vide infra}). Conversely, for partial doubly-excited states with lower single-excitation character, significant contributions from quadruple excitations are expected, making the VTE unsafe when CC4 or CCSDTQ calculations are beyond computational reach. 

Additionally, for both radicals and transition-metal derivatives, strong MR effects and/or non-standard basis set behavior are sometimes observed, which explains why some TBEs were rated as unsafe for these systems, even when CCSDTQ served as the high-level method in Eq.~\eqref{eq1}. Finally, for the largest and least symmetric molecules considered here, we refrained from assigning safe/unsafe labels due to the lack of sufficiently reliable reference data, although most of the reported values are likely within the $\pm 0.050$ eV threshold. Naturally, we cannot guarantee that none of the \textsc{quest} TBEs marked as safe are in fact chemically inaccurate. However, we believe our assessments were rather conservative and that the vast majority of safe TBEs are indeed very close to the true FCI/AVTZ values.

\subsection{Subsets}

In the present work, we have sought to consistently organize the data collected over our seven-year research effort (see Table \ref{Table-1}) into just five distinct subsets. The content of these subsets is summarized in Table \ref{Table-2}, and the associated SI files are structured accordingly. A brief description of each category follows:

\begin{description}
\item[\textsc{main}] This subset includes all results obtained for relatively compact (mostly 1--10 non-H atoms) closed-shell derivatives, encompassing both small inorganic species and a broad array of well-known organic compounds. It consolidates the bulk of our previous work from Refs.~\citenum{Loo18a,Loo20a,Loo20d,Ver21,Loo21a,Loo21,Loo22} and \citenum{Loo24,Kos24,Kny24}, while also incorporating a substantial amount of previously unpublished data, including new molecules and additional ESs.

\item[\textsc{rad}] This is a significant expansion of the dataset from Ref.~\citenum{Loo20d}, focusing on small organic and inorganic radicals. It features additional molecules, new ESs, and, for the first time, a set of quartet ESs. Where comparisons are possible, we also include data from Ref.~\citenum{Son25} in the SI.

\item[\textsc{chrom}] This subset contains ES data for large, closed-shell organic chromophores with 10--16 non-H atoms, such as azobenzene, BODIPY, and naphthalimide. It extends the work presented in Ref.~\citenum{Kny24}, with a strong emphasis on valence transitions. For these sizable systems, reference values are typically based on CC3 with CCSDT/6-31+G(d) corrections for 37 singlet ESs. Due to the complexity of these systems and the use of mixed basis-set corrections [including CCSD and CCSDR(3)], we chose not to label the proposed TBEs as either ``safe'' or ``unsafe''. Nevertheless, we note that for the vast majority of ESs in this set, the computed $\%T_1$ exceeds 85\%, indicating that the TBEs are likely to be sufficiently accurate for benchmarking purposes.

\item[\textsc{bio}] This subset contains previously unpublished data for the five canonical nucleobases (adenine, cytosine, guanine, thymine, and uracil), with TBEs determined using the same methodology as in \textsc{chrom}.

\item[\textsc{tm}] This subset provides the results from Ref.~\citenum{Jac23a} for 11 diatomic species containing a transition metal element (Cu, Sc, Ti, or Zn), covering both closed- and open-shell configurations. Several of these compounds, such as CuH, pose significant challenges for ES theories. Given their distinct behavior compared to traditional organic systems, it is appropriate to treat them as a separate subset. {These results are included in the present SI for completeness, but are not further discussed here, since the \textsc{tm} subset is identical to the one of Ref.}~\citenum{Jac23a}.
\end{description}

As shown in Table \ref{Table-2}, there is a reasonably good balance between valence and Rydberg ESs, singlet and triplet states, as well as reference values obtained for molecules of varying sizes. Further details are provided in Tables S1 and S2 of the SI, where one can observe that most Rydberg ESs are singlet or doublet in nature, whereas transitions in larger molecules are almost exclusively of valence character. While these specificities should be kept in mind when interpreting the statistical results obtained with \textsc{quest}, they also reflect the underlying chemistry: in large $\pi$-conjugated dyes, the lowest ESs are indeed predominantly of valence character. All corresponding data are available in the associated \textsc{GitHub} repository at the following URL: \url{https://github.com/pfloos/QUESTDB}.

\subsection{Benchmarked theories}

For closed-shell molecules, we assessed the performance of the following SR approaches: CIS(D),\cite{Hea94,Hea95} CC2,\cite{Chr95,Hat00} EOM-MP2,\cite{Sta95c} similarity-transformed equation-of-motion-CCSD (STEOM-CCSD),\cite{Noo97,Sou14,Dut18} CCSD,\cite{Pur82,Scu87,Koc90b,Sta93,Sta93b} CCSD(T)(a)*,\cite{Mat16} CCSDR(3),\cite{Chr96b} CCSDT-3,\cite{Wat96,Pro10} CC3,\cite{Chr95b,Koc95,Koc97} CCSDT,\cite{Nog87,Scu88,Kuc01,Kow01,Kow01b} ADC(2),\cite{Tro97,Dre15} ADC(3),\cite{Tro02,Har14,Dre15} and ADC(2.5).\cite{Loo20b} The ADC(2.5) model estimates VTEs as the simple average of ADC(2) and ADC(3) results, i.e., a 50/50 combination shown to approximate the optimal mixing ratio.\cite{Bau22} Naturally, CC3 and CCSDT are benchmarked only when they are not used as the reference high-level methods in Eq.~\eqref{eq1}.  It is worth noting that we do not further discuss the distinctions between linear-response (LR) and EOM formulations at the CC level, nor between LR and intermediate-state representation (ISR) formalisms at the ADC level, as these yield identical VTEs, despite producing different results for other properties.

We also examined the performance of spin-scaled models, including SOS-CC2 and SCS-CC2, \cite{Hel08} as well as SOS-ADC(2) \cite{Hel08,Kra13} with two different parameter sets (\textit{vide infra}).

For open-shell molecules, we also evaluated several SR approaches, although the selection was inherently limited by implementation availability. We assessed the performance of ADC(2), ADC(3), CC2, CCSD, and CC3, consistently considering both U and RO formalisms.

State-of-the-art MR approaches, such as CASPT2, NEVPT2 and CASPT3, \cite{And90,And92,Wer96,Ang01,Ang01b,Ang02} were extensively used in order to assess the accuracy of these methods and to help assigning the nature of the states based on the reference CASSCF wavefunctions. For some troublesome cases involving strong state-mixing, we used multistate variants of CASPT2, such as MS- and XMS-CASPT2. \cite{Fin98,Shi11b}

Given the large volume of data, we naturally employed standard statistical estimators in our analysis: mean signed error (MSE), mean absolute error (MAE), standard deviation of the errors (SDE), root mean squared error (RMSE), {maximal positive and negative errors [Max(+) and Max(-)], as well as the percentage of chemically accurate results (i.e., deviations within $\pm0.05$ eV of the TBE), denoted as $\%$CA.} Readers interested in additional metrics can readily extract them from the files provided in the SI or the \textsc{GitHub} repository.

\subsection{Computational details}

Unless otherwise specified, all \textsc{quest} values were obtained within the frozen-core (FC) approximation, except for beryllium, using large-core treatments for third-row atoms and transition metals, e.g., we freeze 10 electrons in sulfur and phosphorous. {For the \textsc{tm} subset, large cores with frozen $3s$ and $3p$ are applied for the metal centres, yet comparisons with small cores can be found in Ref.} \citenum{Jac23a}.  For all compounds, ground-state geometries (and, in some cases, ES geometries) were optimized at high levels of theory, typically CC3 \cite{Chr95b,Koc95,Koc97} or CCSD(T), \cite{Rag79} with triple-$\zeta$ atomic basis sets. The specific methods used, along with the corresponding Cartesian coordinates, are provided for each compound in the SI.

As mentioned above, we employed a series of increasingly large atomic basis sets incorporating both polarization and diffuse functions: 6-31+G(d), aug-cc-pVDZ (AVDZ), aug-cc-pVTZ (AVTZ), and aug-cc-pVQZ (AVQZ). For some of the smallest systems, even larger basis sets were used to obtain highly accurate CC3 estimates; these results are also given in the SI. All benchmark comparisons were conducted using AVTZ, the basis set for which all TBEs are consistently available.

To determine the TBEs and conduct our benchmarks for closed-shell derivatives, we employed a broad selection of quantum chemistry software packages. \textsc{Q-chem} (versions 5.4, 6.0, and 6.2) \cite{Epi21} was used for EOM-MP2, ADC(2), SOS-ADC(2), and ADC(3) calculations, with the resolution-of-the-identity (RI) approximation systematically applied using the appropriate auxiliary basis sets. \textsc{Turbomole} (versions 7.3, 7.5, and 7.8) \cite{Turbomole,Bal20} was employed to compute CIS(D), CC2, SOS-ADC(2), SOS-CC2, and SCS-CC2 VTEs, also consistently using the RI approach. The default spin-scaling parameters were adopted for the SOS and SCS models. Notably, the SOS-ADC(2) scaling parameters differ between \textsc{q-chem} and \textsc{turbomole}; we therefore label the respective results as [QC] \cite{Kra13} and [TM] \cite{Hel08} throughout the manuscript. \textsc{Orca} (versions 4.2, 5.4, and 6.0) \cite{Nee20,Nee25,orca} was used for all STEOM-CCSD calculations, with only those cases exhibiting active character greater than 98\% being retained. \textsc{Gaussian16} \cite{Gaussian16} and \textsc{dalton} (versions 2016, 2018, and 2020) \cite{dalton} provided the CCSD results, with \textsc{dalton} also used to perform CCSDR(3) calculations. The CCSD(T)(a)$^\star$ and CCSDT-3 transition energies were obtained using \textsc{cfour} (version 2.1). \cite{cfour,Mat20} CC3 calculations were carried out with both \textsc{dalton} and \textsc{cfour}. Finally, CCSDT and CCSDTQ calculations were performed using \textsc{cfour} and \textsc{mrcc} (versions 2017 and 2020), \cite{Rol13,mrcc} with{ \textsc{cfour} and \textsc{mrcc} being also employed for CC4 and CCSDTQP, respectively}.

For open-shell molecules, \textsc{q-chem} was used to perform ADC(2), ADC(3), and CC2 calculations, starting from a U ground-state as well as the RO solution for CC2. RO-CCSD computations were also carried out with this code, while \textsc{gaussian} was employed for the U-CCSD evaluations. Both U- and RO-CC3 VTEs were determined using \textsc{psi4} (versions 1.8 and 3.4). \cite{Psi4} For these open-shell derivatives, \textsc{mrcc} was used to compute VTEs at the CCSDT, CCSDTQ, and CCSDTQP levels.

The SCI calculations were all performed with the CIPSI algorithm \cite{Hur73,Gin13,Gin15,Gar17,Gar18} and carried out with \textsc{quantum package} \cite{Gar19}, following the same protocol as in our previous studies.\cite{Loo18a,Loo19c,Loo20c} The extrapolated FCI estimates are derived from these CIPSI calculations. \cite{Bur24} Extrapolation errors were assessed using the procedure outlined in Ref.~\citenum{Ver21}.

Multiconfigurational calculations were carried out using \textsc{molpro}\cite{Wer20}, following the protocol outlined in Ref.~\citenum{Loo19c}. Additional details, including descriptions of the active spaces for each symmetry representation, are provided in the {\textsc{GitHub} repository} for each system and transition. We performed NEVPT2 calculations in both the partially contracted (PC) and strongly contracted (SC) schemes, as well as CASPT2 and CASPT3 calculations.\cite{Wer96,Bog22} These were done both with and without the IPEA shift.\cite{And93b,And95b,Ghi04} To distinguish between these two cases, we indicate in parentheses whether the IPEA shift is used: CASPT2 includes the shift, while CASPT2(no IPEA) does not. The IPEA shift is set to its default value of $0.25$ hartree. Unless stated otherwise, all CASPT2 and CASPT3 calculations were performed with a level shift of $0.3$ a.u. It is important to note that the Fock operator used in the zeroth-order Hamiltonian was consistently computed using the state-specific density.

We typically employed default or tightened convergence thresholds across all codes. When direct comparisons were possible, such as at the CCSD level, which is implemented in many programs, benchmark tests consistently revealed negligible differences between the various implementations (within $\pm0.001$ eV).

\section{Results and trends}

\subsection{Convergence of the CC series}

Since CC-based methods are primarily used to define the TBEs, it is essential to evaluate their convergence toward the FCI limit. To this end, we begin our analysis with the \textsc{main} subset.

First, let us emphasize that for the tiny compounds for which both well-converged SCI and CCSDTQP energies could be obtained, no inconsistency was observed: the CCSDTQP results fall within the small extrapolation error bars provided by the CIPSI calculations. Next, considering all data in the \textsc{main} subset and all atomic basis sets, we obtained a MAE of 0.005 eV between CCSDTQP and CCSDTQ VTEs (125 cases), with no single absolute error exceeding 0.010 eV when genuine double excitations are excluded. For this latter category (14 cases), the corresponding MAE is, as expected, significantly larger (0.034 eV). Nonetheless, this average difference remains below the 0.050 eV chemical accuracy threshold. In other words, for the vast majority of practical VTE calculations on closed-shell organics, CCSDTQ can be considered nearly exact.

Second, let us examine how the quality of the VTEs degrades when these high-level methods become computationally out of reach. Focusing on all available CCSDTQ VTEs in the \textsc{main} subset (excluding genuine double excitations), we obtain a MSE (MAE) of 0.001 (0.003) eV with CC4, $-0.001$ (0.015) eV with CCSDT, and 0.008 (0.023) eV with CC3. These averages are based on 407, 597, and 597 cases, respectively, and include all available basis sets. In short, CC4 is essentially equivalent to CCSDTQ, while the accuracy gradually decreases when stepping down to CCSDT and CC3. Nevertheless, both methods still yield impressive precision, with average deviations well below the 0.050 eV chemical accuracy threshold. These conclusions are consistent with our earlier CC4 studies on smaller VTE sets. \cite{Loo21b,Loo22} They confirm that CC4, with its $\mathcal{O}(N^9)$ scaling, is a highly reliable method for most ESs.

The picture changes markedly for genuine double excitations. Here, the MSE increases to 0.038 eV with CC4 (32 cases), 0.352 eV with CCSDT (35 cases), and 0.810 eV with CC3 (34 cases), with a systematic overestimation, except for one case (diazete's double excitation with CC4). These results clearly show that accurate treatment of genuine doubles within the CC framework requires inclusion of quadruples. While CCSDT and CC3 can flag the presence of such states, their VTE predictions remain significantly off.

\begin{figure}[htp]
  \includegraphics[width=\linewidth]{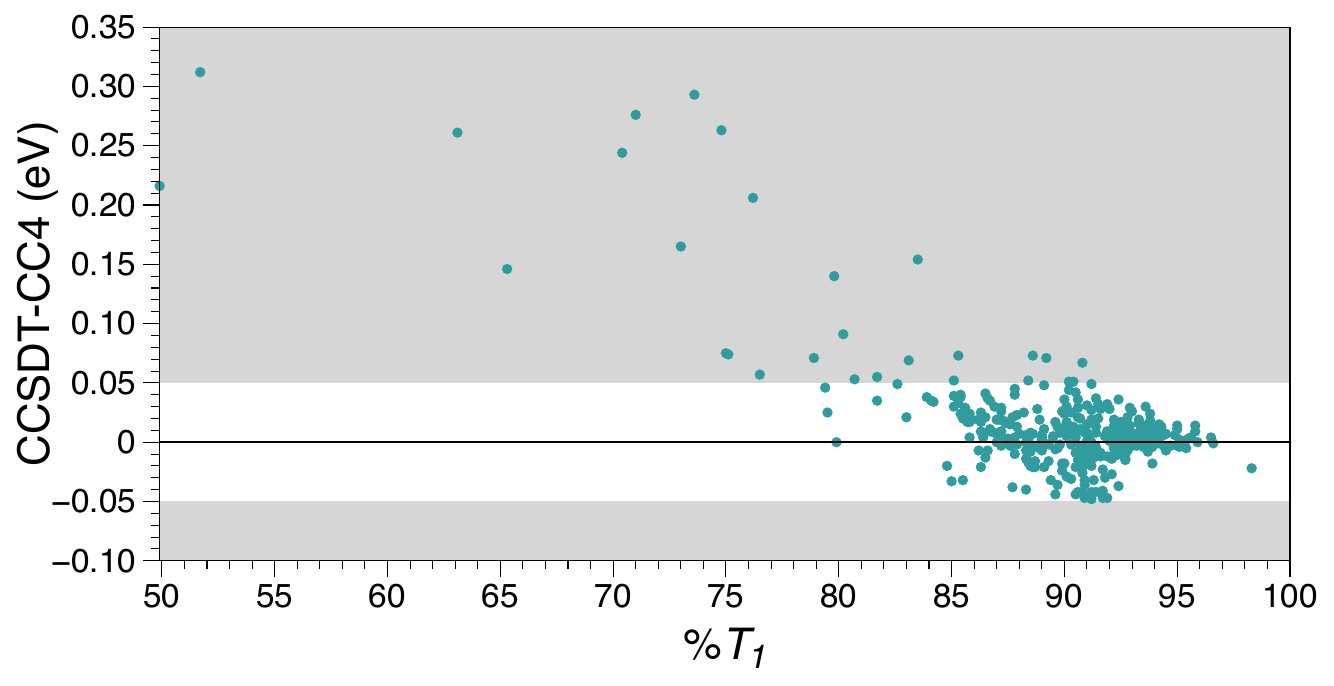}
  \includegraphics[width=\linewidth]{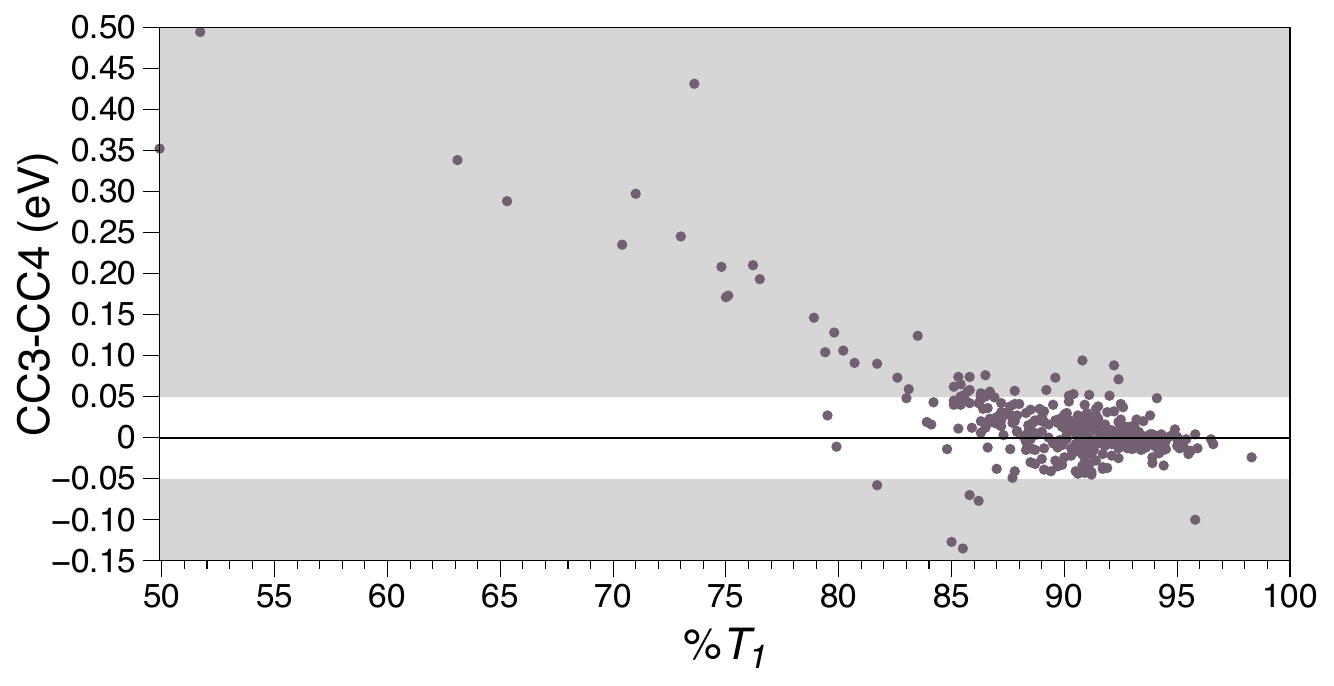}
  \caption{Differences between the CCSDT/6-31+G(d) (top) or CC3/6-31+G(d) (bottom) and CC4/6-31+G(d) VTEs as a function of the single excitation percentage $\%T_1$ for the \textsc{main} subset. The central white zone corresponds to an absolute error smaller than 0.05 eV. Note the difference in vertical axis scale.}
  \label{Fig-c1}
\end{figure}

As illustrated by the genuine double excitations, CC3 and CCSDT errors increase significantly as the single-excitation character of the excited state decreases. To better quantify this effect, we analyzed all 393 CC4/6-31+G(d) VTEs from the \textsc{main} set, and plotted in Figure \ref{Fig-c1} the CC3 and CCSDT errors (using the same basis set and taking CC4 as the reference) as a function of $\%T_1$. As shown, for ESs with $\%T_1 > 85\%$, the CCSDT errors are small and randomly distributed. However, as the double-excitation character increases, the errors become systematically positive, grow roughly linearly, and reach significant values for $\%T_1 < 75\%$. \cite{Kos24} A similar trend is observed for CC3, though with more outliers, faster error growth as $\%T_1$ decreases, and a shift in the onset of significant errors toward slightly higher $\%T_1$ values compared to CCSDT.

\begin{figure}[htp]
  \includegraphics[width=\linewidth]{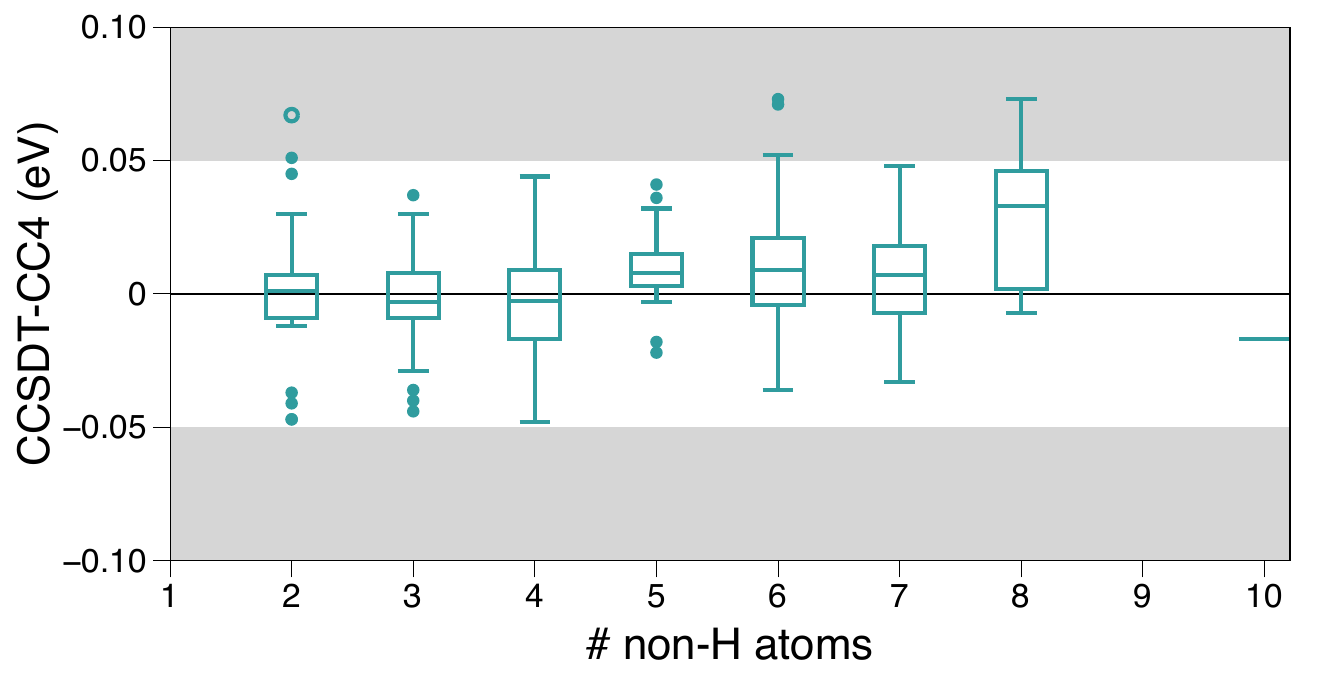}
  \includegraphics[width=\linewidth]{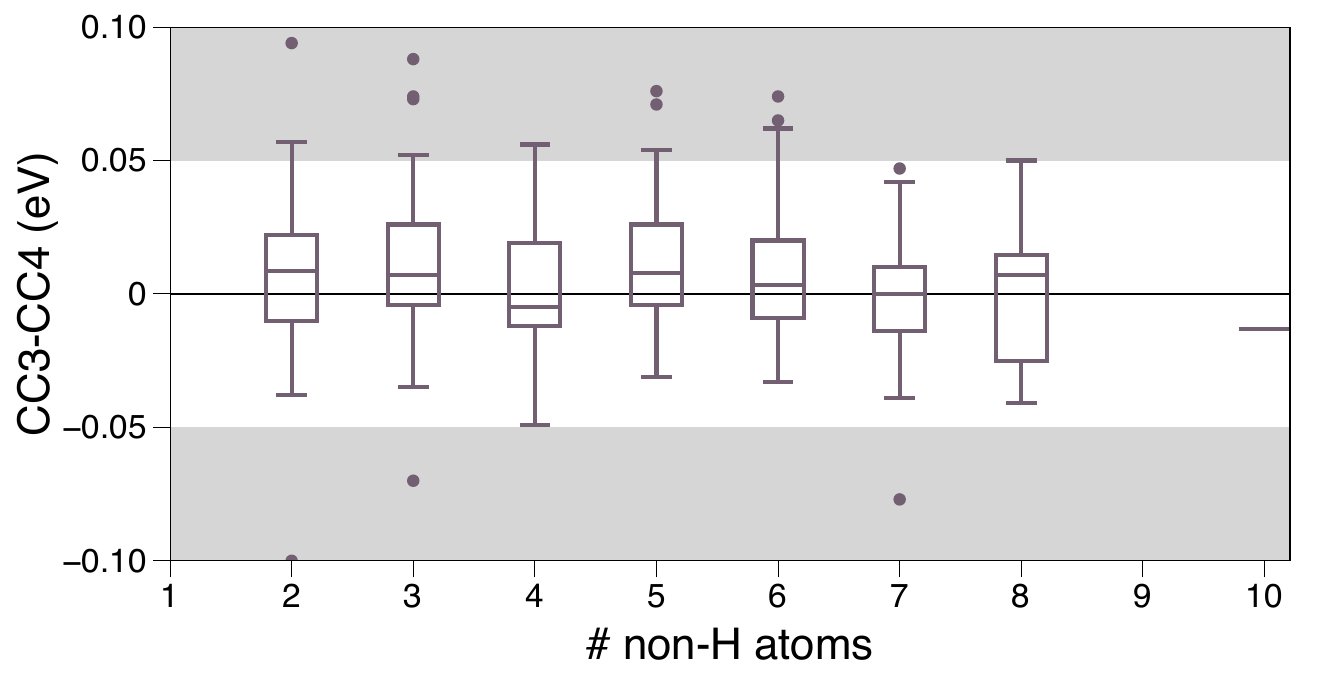}
  \caption{IQR plots of the differences between the CCSDT/6-31+G(d) (top) or CC3/6-31+G(d) (bottom) and CC4/6-31+G(d) VTEs as a function of the molecular size for the ESs of the \textsc{main} subset characterized by $\%T_1 \geq  85\%$. The central white zone corresponds to an absolute error smaller than 0.05 eV. {Standard IQR conventions are followed: the box spans the first to third quartiles with the median marked inside, whiskers extend to 1.5 times the interquartile range, and outliers are shown as individual points.}}
  \label{Fig-c2}
\end{figure}

Nevertheless, one might wonder whether the previously reported CC3 and CCSDT MAEs of 0.015 and 0.023 eV, respectively, are not somewhat optimistic, even for well-behaved, singly ESs. Since these statistics are based on CCSDTQ VTEs as references, they inherently reflect results for the smallest molecules in the \textsc{main} subset, introducing a bias toward few-electron systems that are likely easier to treat with CC methods. It is reasonable to anticipate that CC3 and CCSDT errors could grow for larger, more complex molecules. To explore this possibility, we use CC4/6-31+G(d) results as reference values, as such calculations remain tractable for medium-sized systems, and analyze the corresponding errors for all ESs with $\%T_1 \geq 85\%$ in Figure \ref{Fig-c2}. As can be seen, there is no clear trend in the CCSDT nor CC3 errors as a function of molecular size, an encouraging sign. That said, one might reasonably speculate that more pronounced size effects could emerge with larger basis sets and/or even larger molecules, but investigating this remains beyond our current computational capabilities.

Let us now turn to the radicals included in the \textsc{rad} subset. Once again, the differences between CCSDTQP and CCSDTQ VTEs are statistically negligible. Considering all available basis sets and excluding genuine doubly-excited states, we obtained a MAE as small as 0.002 eV over 295 cases, with no deviation exceeding the chemical accuracy threshold. In a given basis set, CCSDTQ essentially provides exact VTEs for the molecules and states in the \textsc{rad} subset. Using the CCSDTQ results as references, we obtained an MSE of 0.029 (0.093) eV and a MAE of 0.032 (0.099) eV for 600 (594) cases with CCSDT (CC3), again excluding genuine doubles and considering all basis sets. For CC3, approximately 50\% (299 out of 596) of the errors exceed the 0.050 eV threshold. It is therefore clear that CC3 is significantly less efficient for radicals than for closed-shell compounds. While the errors become more acceptable with CCSDT, around 25\% (145 out of 600) of VTEs still do not meet chemical accuracy. Beyond the intrinsic challenges brought by radicals, one of the reason for this less accurate results than for the other subsets is the degenerate nature of some ground electronic states (e.g., in CH). In such case, CC needs double excitations to describe reasonably ESs that are not intrinsically genuine doubles (according to the results of MR calculations).  Given these results and the absence of a CC4 implementation for open-shell species, we have therefore performed, at least, CCSDTQ/6-31+G(d) calculations for all transitions in the \textsc{rad} subset.

\subsection{Basis set effects}

\begin{figure}[htp]
  \includegraphics[width=0.8\linewidth]{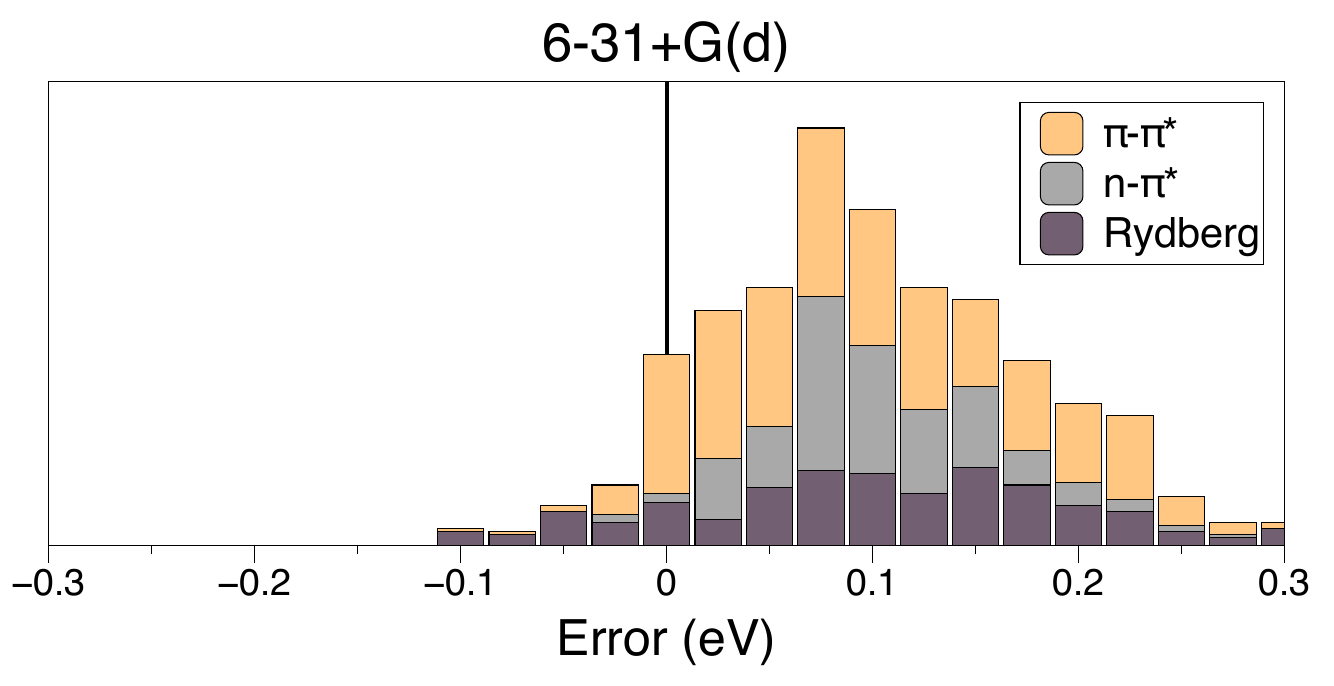}
  \includegraphics[width=0.8\linewidth]{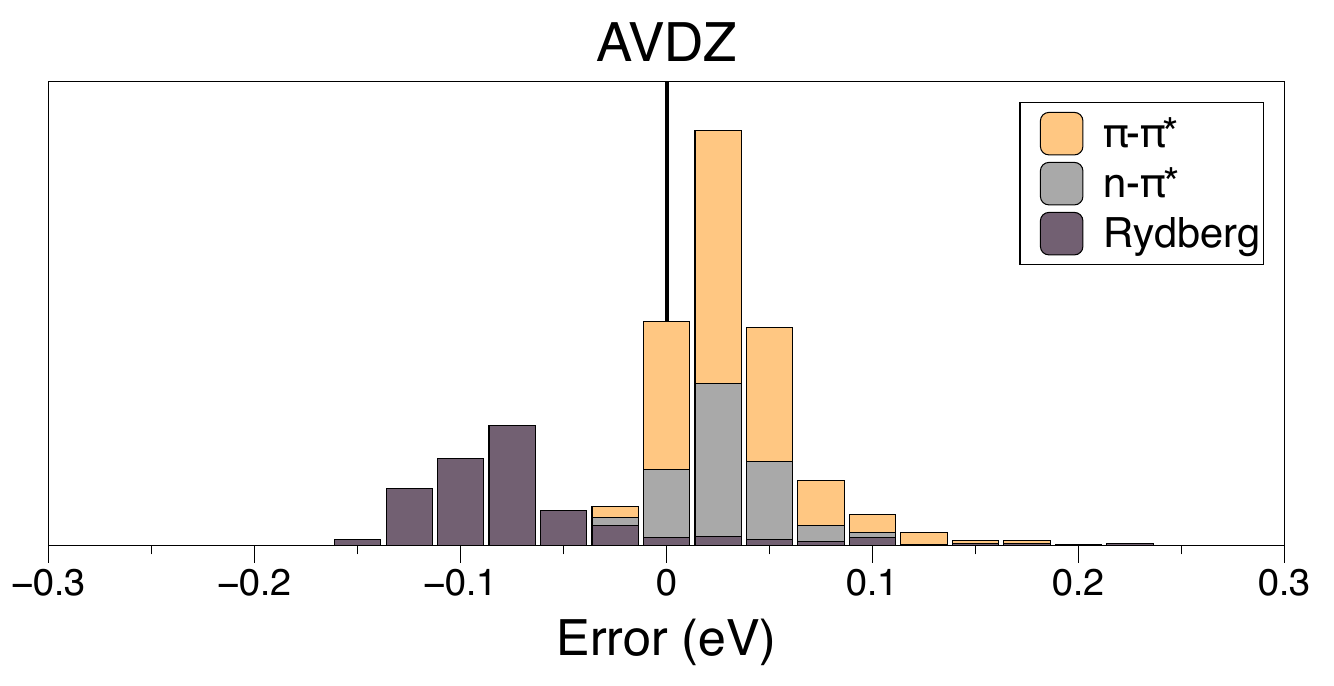}
  \caption{Error distributions relative to AVTZ VTEs obtained with 6-31+G(d) (top, 985 data points) and AVDZ (bottom, 1030 data points) {using stacked histograms}. All calculations were performed at the CC3 level, including all non-genuine double excitations from the \textsc{main}, \textsc{chrom}, and \textsc{bio} subsets.}
  \label{Fig-c3}
\end{figure}

Figure \ref{Fig-c3} presents histograms of the basis set effects on VTEs at the CC3 level, using the AVTZ results as reference and considering all available data from the \textsc{main}, \textsc{chrom}, and \textsc{bio} subsets (excluding genuine doubles). It is evident that 6-31+G(d) almost systematically overestimates the VTEs (MSE: 0.113 eV), with similar mean shifts observed for valence (MSE: 0.101 eV) and Rydberg (MSE: 0.144 eV) excitations. A drastically different picture emerges with AVDZ: although it still overshoots the AVTZ VTEs for valence ESs, the magnitude of the errors is much smaller (MSE: 0.032 eV), while it tends to underestimate (i.e., overstabilize) the Rydberg ESs (MSE: $-0.063$ eV). As a result, the overall average deviation with AVDZ is nearly null (MSE: 0.007 eV), yet the absolute error remains non-negligible (MAE: 0.048 eV). As shown in Section S2 of the SI, differences in errors across molecular sizes and spin states are less pronounced. For example, although triplet ESs and larger molecules appear to yield slightly smaller errors, this stems primarily from the underrepresentation of Rydberg transitions in those categories (\emph{vide supra}).

\begin{figure}[htp]
  \includegraphics[width=0.8\linewidth]{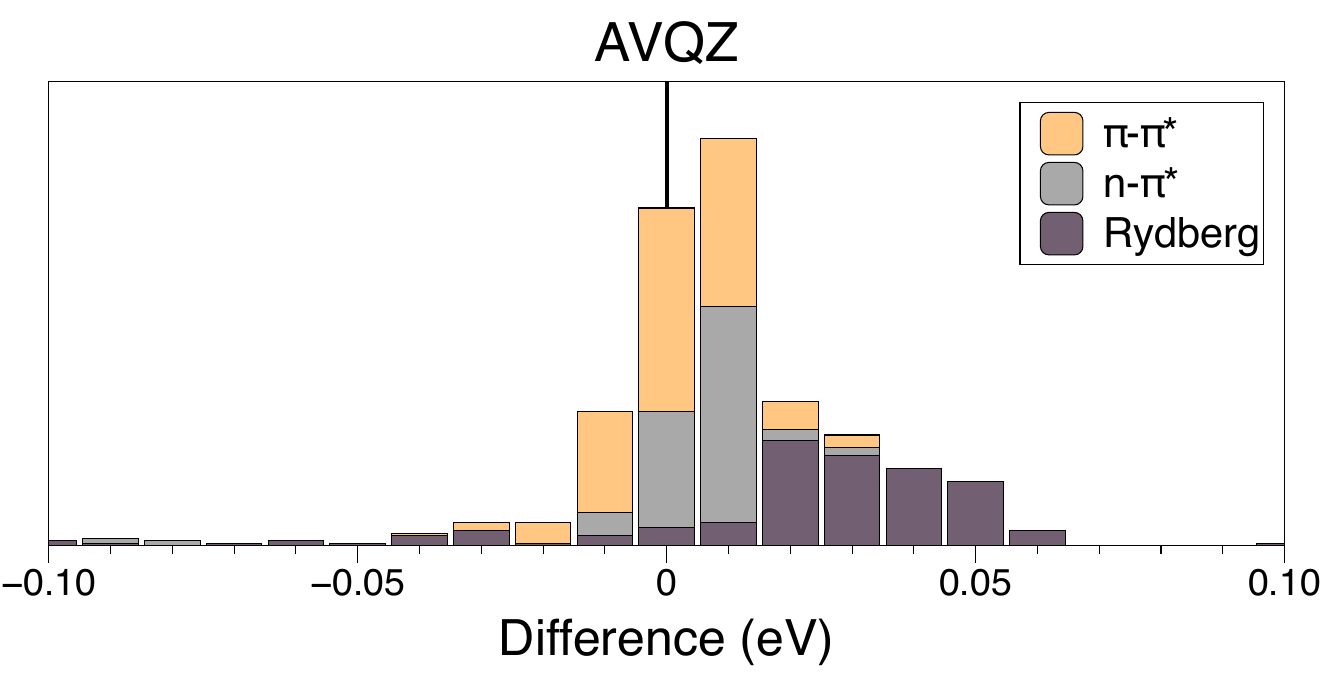}
  \caption{Distribution of deviations from AVTZ VTEs obtained with AVQZ (580 data points) {using stacked histograms}. All calculations were performed at the CC3 level, considering all non-genuine double excitations of the \textsc{main} subset. Note the different horizontal axis scale compared to Figure \ref{Fig-c3}.}
  \label{Fig-c4}
\end{figure}

We now compare AVTZ and AVQZ VTEs, with the distribution of their differences shown in Figure \ref{Fig-c4}. As evidenced, the variations are significantly smaller than those observed in Figure \ref{Fig-c3}, highlighting the already high accuracy of AVTZ-based predictions. For valence ESs, the transition from triple-$\zeta$ to quadruple-$\zeta$ basis sets induces only a negligible positive shift (MSE: 0.002 eV, MAE: 0.009 eV), which likely does not justify the increase in computational cost. Interestingly, Rydberg transitions also exhibit a slight average upshift (MSE: 0.017 eV, MAE: 0.037 eV), suggesting that AVTZ tends to slightly underestimate their energies, and reinforcing the observation that AVDZ errors for Rydberg states are indeed significant. Figure \ref{Fig-c4} also reveals a subset of Rydberg transitions showing substantial negative AVQZ corrections; these typically correspond to highly diffuse Rydberg states found in very compact molecules. For a subset of 296 excited states, CC3/AV5Z calculations were also feasible, and the resulting differences relative to AVQZ are even smaller, MAEs of 0.005 eV for valence and 0.026 eV for Rydberg transitions. For some Rydberg states, additional diffuse functions are likely required to reach basis set convergence. We explored this using d-aug-cc-pVXZ basis sets on the smallest \textsc{quest} molecules, \cite{Loo18a,Chr21} but these basis sets often yield exacerbated state-mixing problems even for relatively small systems. Therefore, we refrain from detailing those results here, though selected values obtained with multiply-augmented basis sets are available in the raw data provided in the SI.
 
\begin{figure}[htp]
  \includegraphics[width=0.8\linewidth]{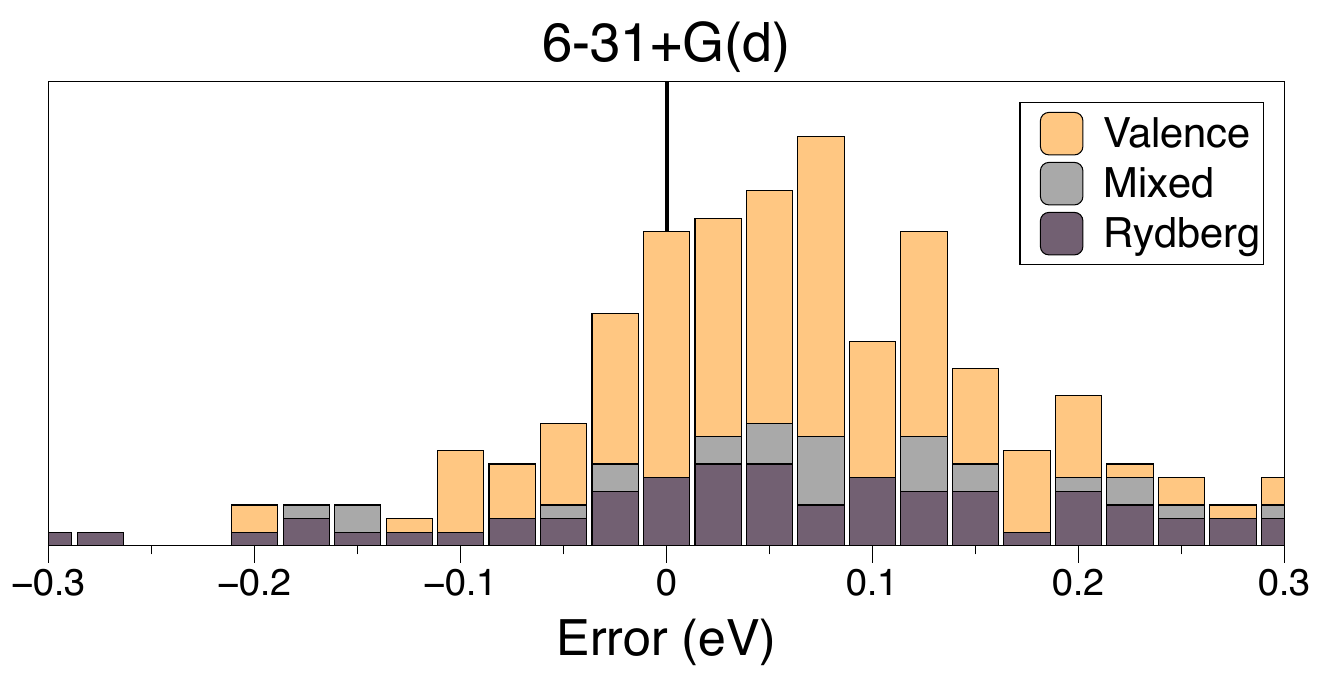}
  \includegraphics[width=0.8\linewidth]{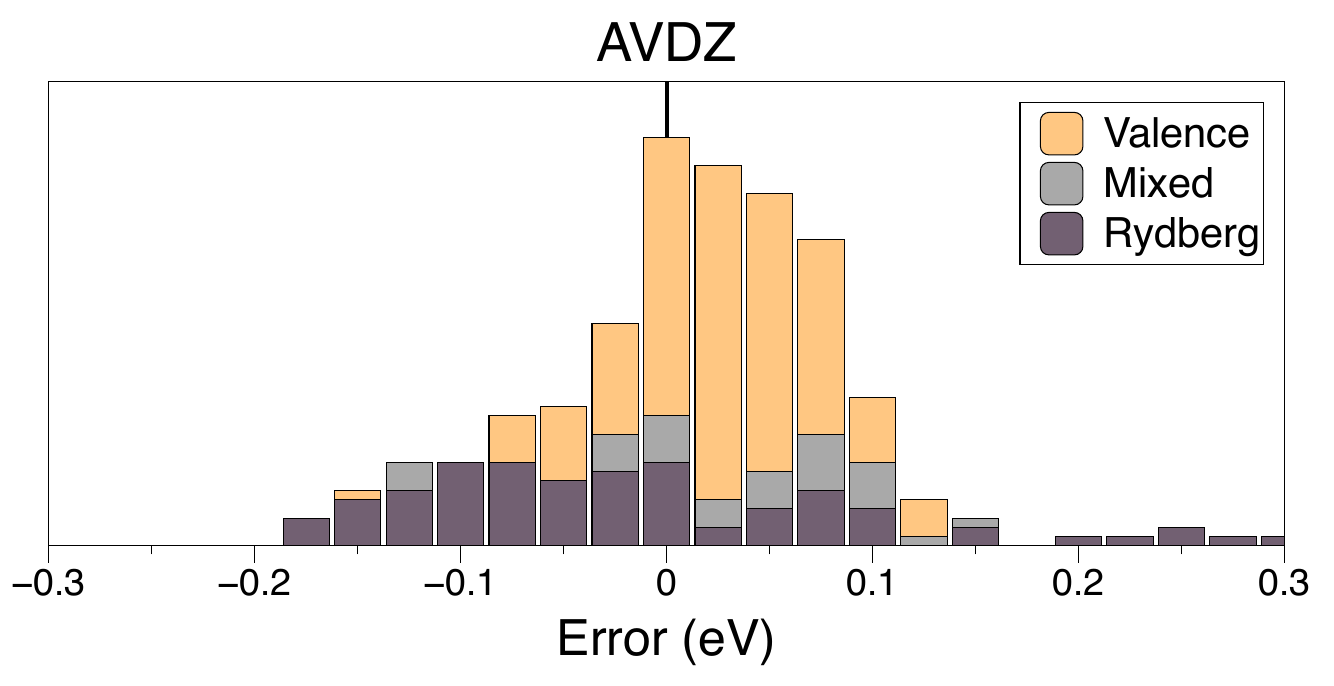}
  \includegraphics[width=0.8\linewidth]{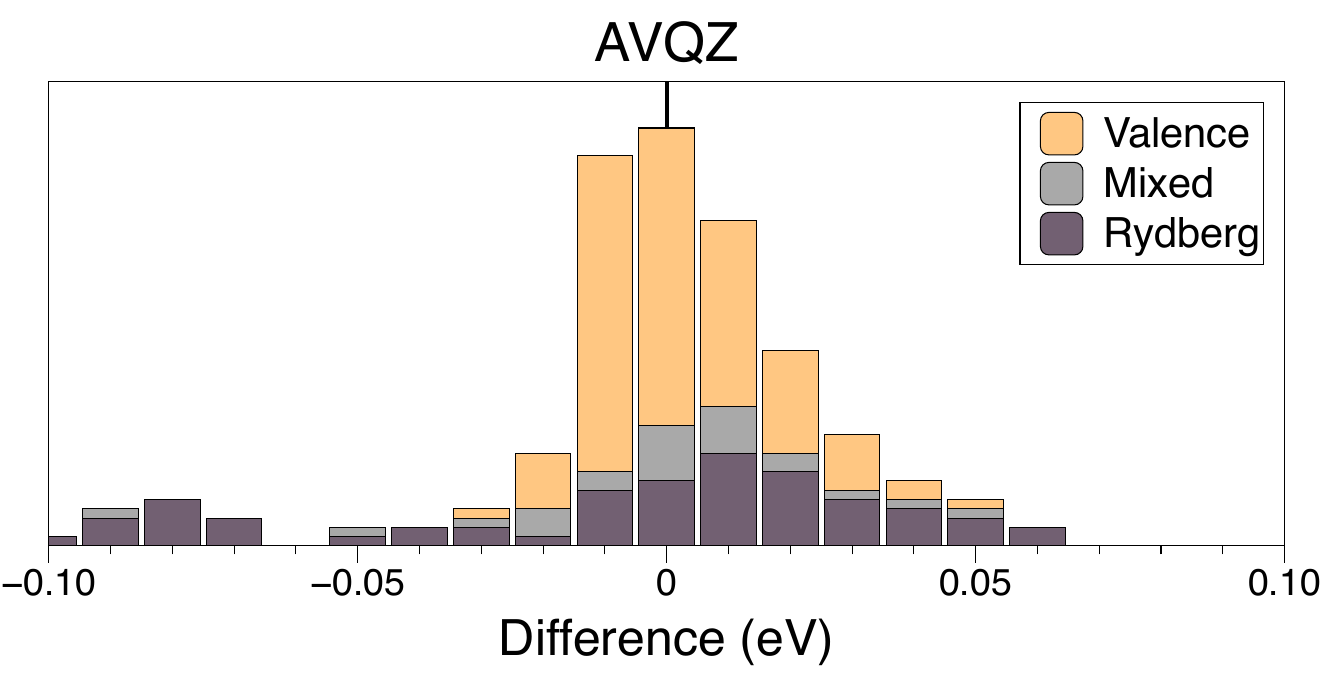}
  \caption{Distribution of deviations from AVTZ VTEs using 6-31+G(d) (top, 259 data points), AVDZ (middle, 267 data points), and AVQZ (bottom, 203 data points) {using stacked histograms}. All calculations were performed at the CC3 level, considering all non-genuine double excitations of the \textsc{rad} subset. Note the different horizontal axis scale in the AVQZ panel.}
  \label{Fig-c5}
\end{figure}

Figure \ref{Fig-c5} presents the analysis of basis set effects for the open-shell molecules in the \textsc{rad} subset. Additional details can be found in Section S2 of the SI, where it is shown that the absolute average impact of basis set variation is comparable for both doublet and quartet states. Overall, the trends mirror those observed for closed-shell systems, though with generally larger basis set effects and a broader spread of values, as expected for open-shell species. For valence transitions, 6-31+G(d) systematically overestimates the VTEs (MSE: 0.060 eV, MAE: 0.087 eV), while AVDZ provides a noticeable improvement (MSE: 0.028 eV, MAE: 0.042 eV). The switch from AVTZ to AVQZ induces only a minor correction for these transitions (MSE: $0.001$ eV, MAE: 0.011 eV), though this shift is very slightly more pronounced than for closed-shell analogs. In the case of Rydberg and mixed transitions, 6-31+G(d) leads to substantial overestimations (MSE: 0.128 eV, MAE: 0.180 eV), which are significantly reduced using AVDZ (MSE: $-0.08$ eV, MAE: 0.093 eV), albeit with a larger dispersion than in closed-shell systems. Switching to AVQZ from AVTZ yields a small average downshift, with mean variations within the chemical accuracy window (MSE: $-0.014$ eV, MAE: 0.038 eV).

\subsection{Validity of the additive approach}

\textsc{Quest} extensively relies on additive basis set corrections to define the TBEs [see Eq.~\eqref{eq1}]. We previously assessed the reliability of this scheme for a smaller set of ESs, \cite{Loo22} and Table \ref{Table-3} presents a broader evaluation based on all available data in the \textsc{main} subset. As shown, applying Eq.~\eqref{eq1} typically reduces the error by approximately a factor of four compared to directly using the corresponding low-order method without correction. In all cases, both MSE and MAE consistently decrease relative to uncorrected results. Notably, the errors are smaller when (i) the small basis set is AVDZ rather than 6-31+G(d), and (ii) the low-order method is of higher quality. Remarkably, starting from CC4/6-31+G(d) and correcting basis set effects with either CCSDT or CC3 yields results extremely close to the actual CC4/AVTZ VTEs (with MAEs around 0.005 eV), which, as demonstrated above, are impressively accurate. These findings echo those reported in Ref.~\citenum{Loo22} and align well with standard ``computational chemistry'' intuition.

\begin{table}[htp]
\caption{\footnotesize Mean signed error (MSE) and mean absolute error (MAE), in eV, obtained by applying Eq.~\eqref{eq1}, {$\Delta {E}_{\text{Large }}^{\text{TBE}}  =  \Delta E_{\text{Small }}^{\text{High}}  +  \qty[ \Delta E_{\text{Large }}^{\text{Low}} 
	- \Delta E_{\text{Small }}^{\text{Low}}  ]$}, for the \textsc{main} subset (excluding genuine double excitations), using AVTZ as the large basis set in all cases. ``\#'' indicates the number of data points. ``None'' refers to the absence of a basis set correction, i.e., it corresponds to the raw error of the low-level method relative to the high-level AVTZ reference.}
\label{Table-3}
\vspace{-0.3 cm}
\footnotesize
\begin{tabular}{lllccc}
\hline
High			& Low		&Small		&\#		&MSE	&MAE\\
\hline
CCSDTQ		&CC4		&AVDZ		&83		&0.000	&0.001\\
			&			&6-31+G(d)	&56		&0.000	&0.002\\
			&			&None		&83		&0.001	&0.003\\
			&CCSDT		&AVDZ		&116		&0.003	&0.004\\
			&			&6-31+G(d)	&71		&-0.001	&0.003\\
			&			&None		&116		&0.002	&0.017\\
			&CC3		&AVDZ		&116		&-0.001	&0.004\\
			&			&6-31+G(d)	&71		&-0.002	&0.007\\
			&			&None		&116		&0.010	&0.026\\			
CC4			&CCSDT		&AVDZ		&124		&0.004	&0.004\\
			&			&6-31+G(d)	&94		&0.000	&0.004\\
			&			&None		&124		&0.004	&0.016\\
			&CC3		&AVDZ		&124		&0.000	&0.004\\
			&			&6-31+G(d)	&94		&-0.003	&0.006\\
			&			&None		&124		&0.008	&0.024\\
CCSDT		&CC3		&AVDZ		&503		&-0.004	&0.005\\
			&			&6-31+G(d)	&457		&-0.003	&0.007\\
			&			&None		&503		&0.004	&0.017\\
\hline
\end{tabular}
\vspace{-0.3 cm}
\end{table}

Table \ref{Table-4} examines the performance of additive basis set corrections for the open-shell species included in the \textsc{rad} subset. In that Table, the target level of theory is CCSDTQ/AVTZ. As observed for closed-shell systems, applying Eq.~\eqref{eq1} leads to a significant reduction in both the mean signed and absolute errors. While, as expected, the errors reported in Table \ref{Table-4} are somewhat larger than those in Table \ref{Table-3}, it is worth noting that the final MAEs remain below 0.010 eV when CCSDT is used as the low-level method. This is particularly reassuring given that CCSDT forms the primary foundation for defining the TBEs of the \textsc{rad} subset.

\begin{table}[htp]
\caption{\footnotesize Mean signed error (MSE) and mean absolute error (MAE), in eV, obtained by applying Eq.~\eqref{eq1} to the \textsc{rad} subset (excluding genuine double excitations), using AVTZ as the large basis set in all cases. See the caption of Table \ref{Table-3} for additional details.}
\label{Table-4}
\vspace{-0.3 cm}
\footnotesize
\begin{tabular}{lllccc}
\hline
High			& Low		&Small		&\#		&MSE	&MAE\\
\hline
CCSDTQ		&CCSDT		&AVDZ		&109		&0.007 	&0.007\\
			&			&6-31+G(d)	&104		&0.007	&0.008\\
			&			&None		&109		&0.023	&0.027\\
			&CC3		&AVDZ		&108		&0.003	&0.008\\
			&			&6-31+G(d)	&103		&0.009	&0.018\\
			&			&None		&109		&0.085	&0.094\\
\hline
\end{tabular}
\vspace{-0.3 cm}
\end{table}

Finally, there is one clear exception to the trends discussed above: genuine double excitations. \cite{Kos24} In such cases, both CC3 and CCSDT tend to incorrectly yield nearly constant transition energies with increasing basis set size. This behavior is unphysical, as larger and more conventional basis set corrections are observed when using higher-level CC methods and/or MR approaches. For this particular class of ESs in non-trivial compounds, we typically relied on the latter methods to determine the basis set correction. Further details on these tests can be found in Ref.~\citenum{Kos24}.

\subsection{Benchmarks}

\subsubsection{Closed-shell molecules}

\begin{table*}[htp]
\caption{\footnotesize Statistical analysis of various theoretical methods based on all TBEs from the \textsc{main}, \textsc{chrom}, and \textsc{bio} subsets, excluding both genuine double excitations and unsafe TBEs. ``\#'' indicates the number of ESs considered; all other values are reported in eV, {except $\%$CA that is given in $\%$}.}
\label{Table-5}
\vspace{-0.3 cm}
\scriptsize
\setlength{\tabcolsep}{1.2pt}
\begin{tabular}{llcccccccccccccccccccccccc}
\hline
&		&\rotatebox{90}{CIS(D)}	&\rotatebox{90}{CC2}	&\rotatebox{90}{EOM-MP2}	&\rotatebox{90}{STEOM-CCSD}	&\rotatebox{90}{CCSD}	&\rotatebox{90}{CCSD(T)(a)*}	&\rotatebox{90}{CCSDR(3)}	&\rotatebox{90}{CCSDT-3}	&\rotatebox{90}{CC3}	&\rotatebox{90}{CCSDT}	&\rotatebox{90}{SOS-ADC(2) [TM]}	&\rotatebox{90}{SOS-CC2}&	\rotatebox{90}{SCS-CC2}	&\rotatebox{90}{SOS-ADC(2) [QC]}	&\rotatebox{90}{ADC(2)}	&\rotatebox{90}{ADC(3)}&	\rotatebox{90}{ADC(2.5)}&	\rotatebox{90}{CASSCF}&	\rotatebox{90}{CASPT2}	&\rotatebox{90}{CASPT2 (no IPEA)}	&\rotatebox{90}{CASPT3}&\rotatebox{90}{CASPT3 (no IPEA)}&	\rotatebox{90}{SC-NEVPT2}	&\rotatebox{90}{PC-NEVPT2}\\
\hline
All		&\#		&999		&1003	&1009	&745		&1009	&599		&614		&597		&886		&458		&1009	&1006	&1007	&1009	&1007	&998		&996		&274		&274		&272		&272		&272		&274		&274\\
		&MSE	&0.14	&0.01	&0.28	&-0.01	&0.12	&0.06	&0.06	&0.05	&0.00	&0.00	&0.19	&0.22	&0.15	&0.01	&-0.03	&-0.07	&-0.05	&0.15	&0.08	&-0.26	&0.11	&0.06	&0.14	&0.10\\
		&MAE	&0.23	&0.15	&0.30	&0.11	&0.14	&0.06	&0.06	&0.06	&0.02	&0.02	&0.21	&0.23	&0.17	&0.13	&0.16	&0.20	&0.08	&0.49	&0.11	&0.28	&0.13	&0.10	&0.16	&0.14\\
		&SDE	&0.25	&0.20	&0.22	&0.12	&0.13	&0.06	&0.05	&0.05	&0.03	&0.02	&0.18	&0.16	&0.15	&0.16	&0.21	&0.22	&0.08	&0.31	&0.08	&0.15	&0.09	&0.07	&0.09	&0.08\\
		&RMSE	&0.30	&0.21	&0.35	&0.15	&0.18	&0.09	&0.08	&0.08	&0.03	&0.03	&0.26	&0.27	&0.21	&0.18	&0.22	&0.25	&0.10	&0.63	&0.17	&0.34	&0.19	&0.15	&0.20	&0.18\\
		&{Max(+)}	&{1.25}	&{0.63}	&{1.32}	&{0.73}	&{1.08}	&{0.59}	&{0.56}	&{0.54}	&{0.20}	&{0.11}	&{1.14}	&{1.06}	&{0.77}	&{0.73}	&{0.63}	&{0.94}	&{0.30}	&{2.17}	&{0.74}	&{0.34}	&{0.93}	&{0.81}	&{0.68}	&{0.50}\\
		&{Max(-)}	&{-1.23}	&{-0.91}	&{-0.48}	&{-0.66}	&{-0.45}	&{-0.18}	&{-0.20}	&{-0.11}	&{-0.13}	&{-0.15}	&{-1.02}	&{-0.39}	&{-1.17}	&{-1.33}	&{-1.36}	&{-0.79}	&{-0.70}	&{-1.20}	&{-0.30}	&{-1.00}	&{-0.29}	&{-0.37}	&{-0.38}	&{-0.60}\\
		&{$\%$CA}	&{15.4}	&{21.4}	&{9.5}	&{29.1}	&{21.1}	&{51.8}	&{52.4}	&{53.8}	&{93.5}	&{92.1}	&{10.3}	&{8.2}	&{14.3}	&{30.5}	&{25.1}	&{15.2}	&{40.0}	&{7.3}	&{36.1}	&{11.0}	&{32.7}	&{43.4}	&{16.8}	&{26.6}\\
&&\\
Singlet	&MSE	&0.09	&-0.04	&0.31	&0.03	&0.17	&0.06	&0.06	&0.05	&0.00	&0.00	&0.18	&0.20	&0.12	&0.00	&-0.07	&-0.03	&-0.05	&0.17	&0.10	&-0.25	&0.15	&0.09	&0.16	&0.11\\
		&MAE	&0.22	&0.16	&0.33	&0.11	&0.17	&0.06	&0.06	&0.06	&0.02	&0.02	&0.21	&0.22	&0.16	&0.14	&0.16	&0.18	&0.08	&0.59	&0.14	&0.27	&0.16	&0.12	&0.18	&0.15\\
Triplet	&MSE	&0.21	&0.07	&0.23	&-0.07	&0.04	&		&		&		&0.01	&-0.01	&0.20	&0.23	&0.18	&0.04	&0.02	&-0.15	&-0.06	&0.12	&0.03	&-0.28	&0.06	&0.01	&0.11	&0.08\\
		&MAE	&0.25	&0.15	&0.25	&0.12	&0.09	&		&		&		&0.01	&0.02	&0.21	&0.24	&0.19	&0.10	&0.15	&0.22	&0.07	&0.34	&0.07	&0.29	&0.07	&0.06	&0.13	&0.11\\
&&\\
Valence	&MSE	&0.22	&0.08	&0.29	&-0.06	&0.13	&0.08	&0.07	&0.07	&0.01	&0.01	&0.19	&0.25	&0.19	&0.00	&0.01	&-0.11	&-0.05	&0.40	&0.07	&-0.32	&0.14	&0.08	&0.14	&0.09\\
		&MAE	&0.24	&0.13	&0.32	&0.12	&0.15	&0.08	&0.08	&0.07	&0.02	&0.02	&0.21	&0.26	&0.20	&0.12	&0.14	&0.21	&0.07	&0.47	&0.11	&0.33	&0.14	&0.11	&0.17	&0.14\\
$\npi$	&MSE	&0.15	&-0.01	&0.29	&-0.01	&0.20	&0.10	&0.09	&0.08	&0.01	&0.01	&0.27	&0.34	&0.22	&0.04	&-0.10	&0.07	&-0.02	&0.44	&-0.03	&-0.44	&0.13	&0.10	&0.11	&0.06\\
		&MAE	&0.20	&0.10	&0.32	&0.09	&0.20	&0.10	&0.10	&0.08	&0.02	&0.02	&0.27	&0.34	&0.23	&0.13	&0.15	&0.17	&0.06	&0.45	&0.08	&0.44	&0.13	&0.11	&0.14	&0.12\\
$\ppi$	&MSE	&0.26	&0.13	&0.32	&-0.09	&0.09	&0.08	&0.07	&0.06	&0.01	&0.01	&0.15	&0.20	&0.18	&-0.02	&0.08	&-0.21	&-0.07	&0.37	&0.13	&-0.24	&0.14	&0.08	&0.17	&0.11\\
		&MAE	&0.28	&0.16	&0.33	&0.14	&0.13	&0.08	&0.07	&0.06	&0.02	&0.02	&0.18	&0.21	&0.18	&0.12	&0.14	&0.23	&0.08	&0.48	&0.14	&0.26	&0.15	&0.11	&0.19	&0.15\\
Rydberg	&MSE	&-0.05	&-0.18	&0.25	&0.09	&0.09	&0.02	&0.02	&0.03	&-0.01	&-0.02	&0.18	&0.13	&0.02	&0.03	&-0.15	&0.02	&-0.07	&-0.52	&0.10	&-0.11	&0.05	&0.01	&0.12	&0.12\\
		&MAE	&0.20	&0.20	&0.26	&0.11	&0.09	&0.03	&0.03	&0.04	&0.02	&0.02	&0.20	&0.15	&0.11	&0.13	&0.20	&0.18	&0.08	&0.55	&0.12	&0.14	&0.07	&0.07	&0.14	&0.14\\
&&\\
Tiny		&MSE	&0.13	&0.05	&0.04	&-0.01	&0.05	&0.02	&0.02	&0.02	&0.00	&0.00	&0.16	&0.18	&0.14	&0.02	&0.02	&-0.15	&-0.06	&		&		&		&		&		&		&\\
		&MAE	&0.29	&0.25	&0.13	&0.10	&0.08	&0.03	&0.03	&0.03	&0.02	&0.02	&0.22	&0.23	&0.22	&0.18	&0.23	&0.23	&0.09	&		&		&		&		&		&		&\\
Small	&MSE	&0.11	&0.02	&0.16	&-0.02	&0.08	&0.03	&0.03	&0.04	&0.01	&0.00	&0.17	&0.22	&0.15	&0.01	&-0.05	&-0.10	&-0.07	&0.02	&0.08	&-0.19	&0.09	&0.04	&0.14	&0.10\\
		&MAE	&0.22	&0.17	&0.19	&0.12	&0.11	&0.04	&0.04	&0.04	&0.02	&0.02	&0.20	&0.22	&0.18	&0.13	&0.17	&0.21	&0.09	&0.47	&0.11	&0.21	&0.10	&0.08	&0.15	&0.13\\
Medium	&MSE	&0.13	&-0.02	&0.36	&0.00	&0.15	&0.08	&0.07	&0.07	&0.00	&0.01	&0.21	&0.22	&0.14	&0.02	&-0.03	&-0.04	&-0.04	&0.41	&0.06	&-0.40	&0.16	&0.12	&0.15	&0.09\\
		&MAE	&0.23	&0.14	&0.36	&0.12	&0.16	&0.08	&0.07	&0.07	&0.02	&0.02	&0.22	&0.22	&0.16	&0.12	&0.14	&0.19	&0.06	&0.53	&0.12	&0.42	&0.17	&0.14	&0.18	&0.15\\
Large	&MSE	&0.19	&0.00	&0.46	&-0.06	&0.15	&0.08	&0.08	&0.06	&-0.01	&		&0.21	&0.23	&0.16	&0.01	&-0.03	&-0.06	&-0.05	&0.50	&0.05	&-0.22	&0.09	&0.05	&-0.14	&-0.17\\
   		&MAE	&0.22	&0.10	&0.46	&0.12	&0.16	&0.08	&0.08	&0.06	&0.02	&		&0.22	&0.24	&0.16	&0.10	&0.12	&0.18	&0.07	&0.50	&0.05	&0.22	&0.09	&0.05	&0.14	&0.17\\
\hline
\end{tabular}
\vspace{-0.3 cm}
\end{table*}

Having established the quality of the \textsc{quest} TBEs, we now turn to benchmarks that rely on these reference values. We chose to consider all TBEs from the \textsc{main}, \textsc{chrom}, and \textsc{bio} subsets together, excluding only genuine double excitations and unsafe transitions (i.e., those with $\%T_1 < 85\%$ in the latter two subsets). It is worth noting that retaining all unsafe transitions, while still excluding genuine doubles, would not drastically alter the statistical results, leading only to a modest increase in the observed errors.

\begin{figure*}[htp]
  \includegraphics[width=0.99\linewidth,viewport=1.7cm 5.0cm 20.0cm 28.5cm,clip]{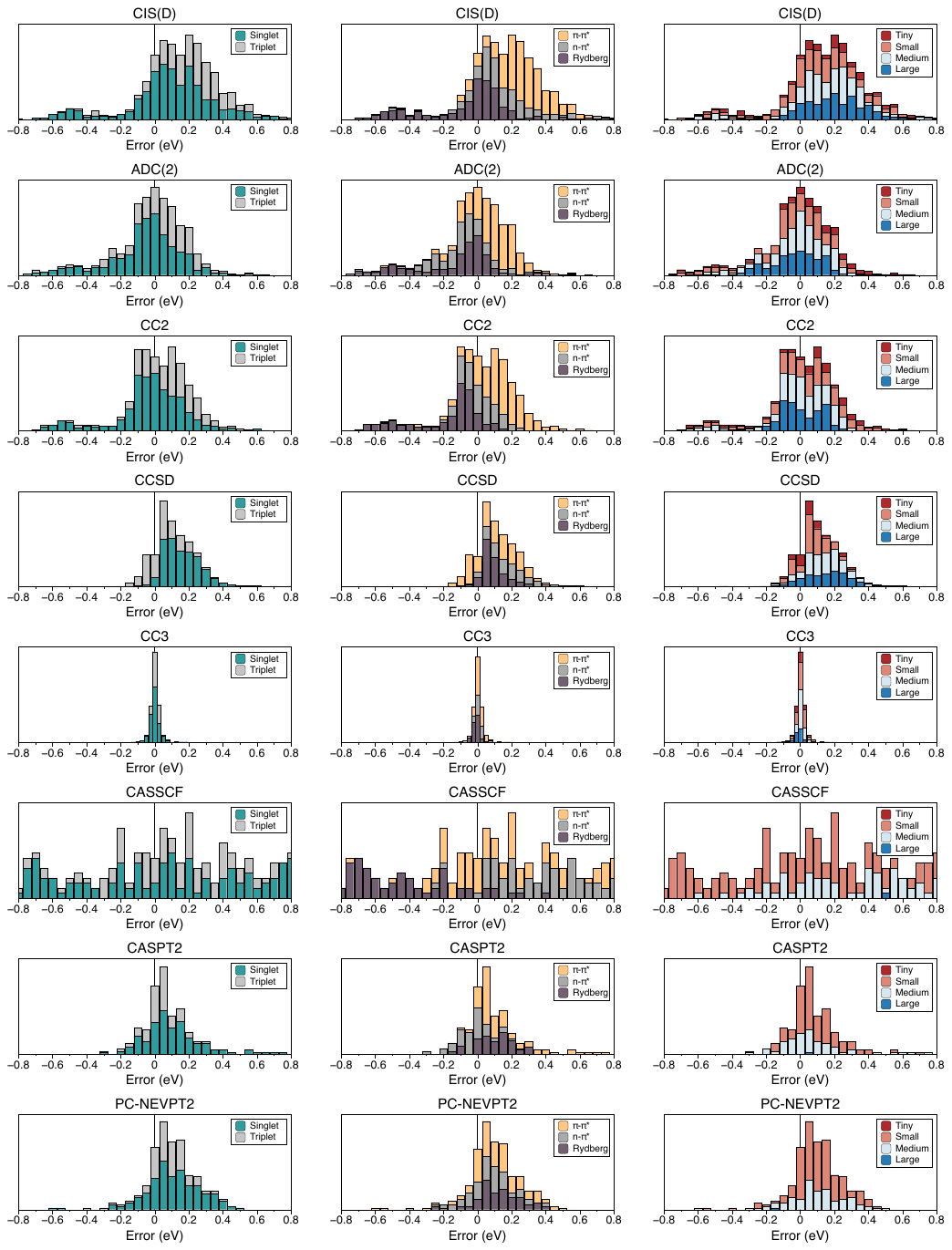}
  \caption{Distribution of the errors on VTEs for various levels of theory  {using stacked histograms and} considering all data from the \textsc{main}, \textsc{chrom}, and \textsc{bio} subsets, excluding genuine double excitations and unsafe TBEs. From top to bottom: CIS(D), ADC(2), CC2, CCSD,  CC3, CASSCF, CASPT2 (with IPEA), and PC-NEVPT2. From left to right: impact of the spin symmetry, effect of ES nature, and influence of molecular size.}
    \label{Fig-c6}
\end{figure*}

The results obtained using these reference VTEs are presented in Table \ref{Table-5} for all evaluated methods. Figure \ref{Fig-c6} displays histograms of the error distributions for several widely used methods (complete plots are available in the SI). We begin by examining the performance of SR methods.

The computationally most affordable method, CIS(D), yields a MAE of 0.23 eV, accompanied by a relatively large error spread, as indicated by an SDE of {0.25} eV, the highest among all SR approaches considered here. This relatively poor performance is consistent with earlier findings. \cite{Goe10a,Jac15b,Kan17,Ver21} On the positive side, the CIS(D) error is largely unaffected by molecular size or by whether singlet or triplet ESs are considered. In contrast, CIS(D) tends to underestimate the VTEs of Rydberg states and to significantly overestimate those of $\ppi$ transitions, as illustrated in the top right panel of Figure \ref{Fig-c6}.

EOM-MP2, while less commonly employed but also computationally inexpensive, consistently overestimates transition energies, with a striking MSE of +0.28 eV, the largest of all methods evaluated. These overestimations are somewhat reduced for triplet compared to singlet states, and for Rydberg versus valence transitions. However, the error increases steadily with molecular size (see Figure S7 in the SI), suggesting that EOM-MP2 is unlikely to be a reliable choice for large systems.

The two most widely used second-order SR methods, ADC(2) and CC2, exhibit very similar error patterns and overall deliver more accurate transition energies than both CIS(D) and EOM-MP2. These trends are consistent with prior studies. \cite{Hat05c,Win13,Har14,Jac15b,Kan17,Ver21} Both ADC(2) and CC2 achieve a MAE of {0.15--0.16} eV with negligible MSE, and both display a slight tendency to overestimate singlet energies and underestimate triplet ones. Encouragingly, the errors associated with these methods tend to decrease with molecular size, and both perform better for valence excitations, especially $\npi$ transitions, than for Rydberg states; these two trends are likely interrelated. The small bump observed around $-0.5$ eV in the error distributions of Figure \ref{Fig-c6} can be attributed primarily to Rydberg transitions. In summary, ADC(2) and CC2 appear well suited for practical simulations of organic molecules, {although they achieve chemical accuracy only 20--25$\%$ of the time.}

Spin-scaling variants of ADC(2) and CC2 can be employed at the same (or even reduced) computational cost compared to the standard methods. Among them, SCS-CC2 generally performs better than SOS-CC2, notably reducing the error spread (SDE of 0.15 eV \textit{vs} 0.20 eV for CC2) and yielding nearly uniform MAEs across all excitation families, except for Rydberg transitions. However, this improvement comes at the cost of a somewhat systematic overestimation of the transition energies. For SOS-ADC(2), two parameter sets were tested, with the \textsc{q-chem} parameters proving more effective, unsurprisingly, as they were specifically optimized for transition energies of organic molecules. \cite{Kra13} SOS-ADC(2)[QC] achieves MAEs below 0.20 eV across all subgroups considered, with an overall MAE of 0.13 eV and an SDE of {0.16} eV, an excellent performance given its relatively low computational cost. These findings regarding the benefits of spin-scaling align well with earlier studies. \cite{Goe10a,Jac15b,Taj20a,Ver21}

Overall, CCSD delivers a slightly lower MAE than CC2 (0.14 eV \textit{vs} {0.15} eV), along with a notably smaller error spread (SDE of 0.13 eV \textit{vs} 0.20 eV). However, as illustrated in Figure \ref{Fig-c6}, the CCSD error patterns are clearly biased: triplet states are significantly more accurate than singlet excitations, Rydberg transitions outperform valence ones, and, critically, the deviations from the TBEs tend to increase with molecular size. In this regard, CC2 shows a clear advantage for the largest systems examined in this work. While the overestimation trend of CCSD is well documented in the literature, \cite{Sch08,Car10,Wat13,Sou14,Kan14,Kan14b,Kan17,Ver21} the present benchmark reveals rather substantial deviations (with a MAE of 0.16 eV for medium and large compounds). Accordingly, CCSD may not be well-suited for assessing the performance of CC2 or ADC(2) for realistic organic systems. {ADC(2), CC2 and CCSD also deliver similar percentages of chemical accuracy}. For instance for triangulenes with inverted singlet-triplet gaps, the second-order methods proved more accurate. \cite{Loo23a,Loo23ac}

The same disappointing scenario, i.e., higher computational cost without clear gains in accuracy, emerges when comparing ADC(2) and ADC(3). Overall, ADC(3) delivers less accurate results than its second-order counterpart, except for Rydberg excitations, although the MAE remains relatively consistent across all subgroups considered. In contrast, the ADC(2.5) scheme, defined as the simple average of ADC(2) and ADC(3) excitation energies, \cite{Loo20b} yields substantial improvements. As shown in Table \ref{Table-5}, with ADC(2.5), all MSEs and MAEs are within 0.10 eV, with no marked bias for any specific subgroup aside from a general underestimation trend (MSE of $-0.05$ eV). {ADC(2.5) attains chemical accuracy for 40$\%$ of the cases.} These findings are consistent with recent studies focused on smaller sets of transitions. \cite{Loo20b,Ver21,Bau22,Loo24}

\begin{table*}[htp]
\caption{\footnotesize Statistical analysis of various theoretical methods based on all TBEs from the \textsc{rad} subset, excluding both genuine double excitations and unsafe TBEs. See caption of Table \ref{Table-5} for more details.}
\label{Table-6}
\vspace{-0.3 cm}
\scriptsize
\begin{tabular}{llccccccccccccccc}
\hline
&		&\rotatebox{90}{U-CC2}	&\rotatebox{90}{RO-CC2}	&\rotatebox{90}{U-CCSD}	&\rotatebox{90}{RO-CCSD}	&\rotatebox{90}{U-CC3}	&\rotatebox{90}{RO-CC3}	&	\rotatebox{90}{U-ADC(2)}	&\rotatebox{90}{U-ADC(3)}		&\rotatebox{90}{CASSCF}&	\rotatebox{90}{CASPT2}	&	\rotatebox{90}{CASPT2  (no IPEA)}&\rotatebox{90}{CASPT3} &\rotatebox{90}{CASPT3  (no IPEA)}&\rotatebox{90}{SC-NEVPT2}	&\rotatebox{90}{PC-NEVPT2} \\
\hline
All		&\#			&207		&208		&216		&216		&212		&216		&210		&186		&143		&143		&143		&143		&143		&143		&143		\\
		&MSE		&0.32	&0.28	&0.24	&0.20	&0.08	&0.07	&0.35	&-0.06	&0.02	&0.02	&-0.07	&0.02	&0.00	&0.04	&0.02	\\
		&MAE		&0.35	&0.33	&0.24	&0.21	&0.08	&0.07	&0.37	&0.24	&0.37	&0.08	&0.13	&0.07	&0.06	&0.11	&0.08	\\
		&SDE		&0.32	&0.32	&0.23	&0.22	&0.10	&0.09	&0.33	&0.31	&0.35	&0.09	&0.11	&0.08	&0.07	&0.10	&0.08	\\
		&RMSE		&0.46	&0.45	&0.34	&0.30	&0.13	&0.12	&0.48	&0.38	&0.48	&0.12	&0.17	&0.11	&0.10	&0.14	&0.12	\\
		&{Max(+)}	&{1.89}	&{1.77}	&{1.49}	&{1.37}	&{0.61}	&{0.58}	&{2.23}	&{1.59}	&{1.84}	&{0.46}	&{0.43}	&{0.41}	&{0.38}	&{0.35}	&{0.31}	\\
		&{Max(-)}	&{-0.47}	&{-0.53}	&{-0.13}	&{-0.14}	&{-0.25}	&{-0.26}	&{-0.46}	&{-1.31}	&{-1.36}	&{-0.50}	&{-0.57}	&{-0.46}	&{-0.48}	&{-0.50}	&{-0.56}	\\
		&{$\%$CA}	&{7.2}	&{8.2}	&{19.4}	&{25.0}	&{54.2}	&{60.2}	&{7.6}	&{25.3}	&{9.8}	&{45.5}	&{25.9}	&{55.2}	&{66.4}	&{31.5}	&{45.5}	\\	
		&			&		&		&		&		&		&		&		&		&		&		&		&		&		&		&		\\
Doublet	&MSE		&0.27	&0.22	&0.22	&0.18	&0.06	&0.05	&0.32	&0.00	&0.03	&0.02	&-0.05	&0.03	&0.01	&0.04	&0.02	\\
		&MAE		&0.31	&0.27	&0.22	&0.18	&0.07	&0.06	&0.35	&0.20	&0.38	&0.09	&0.12	&0.07	&0.06	&0.11	&0.08	\\
Quartet	&MSE		&0.54	&0.56	&0.35	&0.31	&0.18	&0.16	&0.48	&-0.31	&-0.02	&-0.02	&-0.18	&0.00	&-0.04	&0.05	&0.02	\\
		&MAE		&0.55	&0.56	&0.35	&0.31	&0.18	&0.16	&0.50	&0.41	&0.32	&0.06	&0.18	&0.05	&0.04	&0.10	&0.06	\\
		&			&		&		&		&		&		&		&		&		&		&		&		&		&		&		&		\\
Valence	&MSE		&0.45	&0.42	&0.30	&0.26	&0.11	&0.09	&0.44	&-0.12	&0.23	&0.04	&-0.08	&0.04	&0.02	&0.09	&0.05	\\
		&MAE		&0.46	&0.43	&0.30	&0.26	&0.11	&0.10	&0.46	&0.34	&0.30	&0.07	&0.13	&0.06	&0.04	&0.10	&0.06	\\
Rydberg	&MSE		&0.11	&0.08	&0.14	&0.12	&0.03	&0.02	&0.21	&0.05	&-0.49	&-0.06	&-0.07	&-0.05	&-0.05	&-0.06	&-0.05	\\
		&MAE		&0.18	&0.17	&0.14	&0.12	&0.05	&0.04	&0.24	&0.09	&0.50	&0.11	&0.14	&0.07	&0.07	&0.10	&0.09	\\
		&			&		&		&		&		&		&		&		&		&		&		&		&		&		&		&		\\
Tiny		&MSE		&0.34	&0.32	&0.23	&0.19	&0.09	&0.07	&0.37	&-0.11	&0.00	&0.01	&-0.09	&0.01	&-0.01	&0.04	&0.01	\\
		&MAE		&0.37	&0.35	&0.23	&0.20	&0.09	&0.08	&0.39	&0.22	&0.36	&0.08	&0.13	&0.07	&0.05	&0.10	&0.07	\\
Small	&MSE		&0.25	&0.16	&0.28	&0.23	&0.05	&0.03	&0.27	&0.14	&0.06	&0.05	&0.01	&0.05	&0.05	&0.07	&0.05	\\
		&MAE		&0.29	&0.23	&0.29	&0.24	&0.05	&0.04	&0.33	&0.31	&0.40	&0.09	&0.15	&0.07	&0.07	&0.11	&0.11	\\
\hline
\end{tabular}
\vspace{-0.3 cm}
\end{table*}

Another reasonably affordable method delivering valuable results is STEOM-CCSD, which yields a MAE of {0.11} eV and a SDE of 0.13 eV. It offers a good balance of accuracy across the various families of ESs, although it tends to underestimate valence transition energies and overestimate those of Rydberg states. This imbalance implies that the relative ordering of transitions may not be optimal. The reliability of STEOM-CCSD was also emphasized in previous benchmarks on Thiel's set. \cite{Sou14,Dut18}

The two CC approaches incorporating perturbative triple excitations, CCSD(T)(a)\textsuperscript{$\star$} and CCSDR(3), which can be regarded as ``CCSD(T) for excited states'', deliver essentially identical results and consistently improve upon CCSD, notably yielding chemically accurate Rydberg excitation energies. Like their parent method, CCSD, their performance however deteriorates for larger molecules, where they exhibit a MAE of approximately 0.08 eV and a nearly systematic overestimation of the VTEs. While this level of accuracy is suitable for most applications, it comes at a significantly higher computational cost compared to CC2, which achieves a comparable MAE of 0.10 eV for large systems. The CCSDT-3 approach, which includes iterative triples, offers only moderate improvements over the perturbative variants and still shows a tendency toward overestimation. Given the modest gain in accuracy, the added computational expense is likely unjustified. Nonetheless, it is worth emphasizing that CCSDT-3 has been found particularly effective for charge-transfer transitions. \cite{Koz20,Loo21a}

CC3 is exceptionally accurate, achieving a MAE of 0.02 eV, a SDE of {0.03} eV, {chemical accuracy in $93.5\%$ of the cases,} with outstanding performance across all types of transitions and no noticeable decline in accuracy as the system size increases. We thus reaffirm the high reliability of CC3, which we have previously highlighted. The improvements offered by CCSDT are modest and, given its higher computational scaling with system size, do not appear to justify its use over CC3. In light of this, recent advancements in methods that enable faster CC3 calculations \cite{Fra20,Pau21,Pau22,Mic24} are certainly welcome.

For the MR approaches, we primarily focused on small and medium-sized molecules, and the results presented here are essentially consistent with those of Refs.~\citenum{Sar22} and \citenum{Bog22}, using carefully designed active spaces (details provided in the SI). All MR methods yield errors that are reasonably similar across the different families of transitions. State-average CASSCF delivers poor results, with a MAE of 0.49 eV and an SDE of 0.31 eV, both the worst values among the methods tested. Clearly, dynamical correlation is necessary to achieve reasonable accuracy in transition energies. CASPT2 (with IPEA corrections) provides much improved results, with a MAE of 0.11 eV and an SDE of 0.08 eV, similar to the ADC(2.5) results, but still significantly worse than CC3 (see Figure \ref{Fig-c6}), for both singlet and triplet transitions. For CASPT2, neglecting the IPEA shift leads to large underestimations of transition energies, particularly for valence transitions (MSE of {$-0.26$} eV), which is expected due to the use of a large atomic basis set. \cite{Zob17} Switching to CASPT3 significantly reduces the impact of the IPEA shift, with the absence of such correction proving globally beneficial. The MAE and SDE for CASPT3 (no IPEA) are 0.10 eV and 0.07 eV, respectively, only slightly better than the CASPT2 results, but at a much higher computational cost. Another option to address the IPEA issue is PC-NEVPT2, which shows trends very similar to CASPT2, though with a slight increase in the overestimation (MSE of 0.10 eV instead of 0.08 eV for CASPT2).

{The above analysis excludes genuine doubles. Considering only the safe genuine doubles in \textsc{main}, we obtain MAE of 0.90 eV (0.45 eV) for CC3 (CCSDT) with systematic overestimation of the TBEs. For these transitions, the MR approaches are, as expected, much more efficient, with MAE of 0.10 eV for CASPT2 (with IPEA), 0.09 eV for CASPT3 (irrespective of the use of IPEA or not), and 0.06 eV for PC-NEVPT2. This clearly illustrates that the methodological effects differ for genuine doubles and other excited states. A more detailed discussion can be found in Ref.} \citenum{Kos24}.

\subsubsection{Open-shell molecules}

\begin{figure*}[htp]
  \includegraphics[width=0.99\linewidth,viewport=1.7cm 5.0cm 20.0cm 28.5cm,clip]{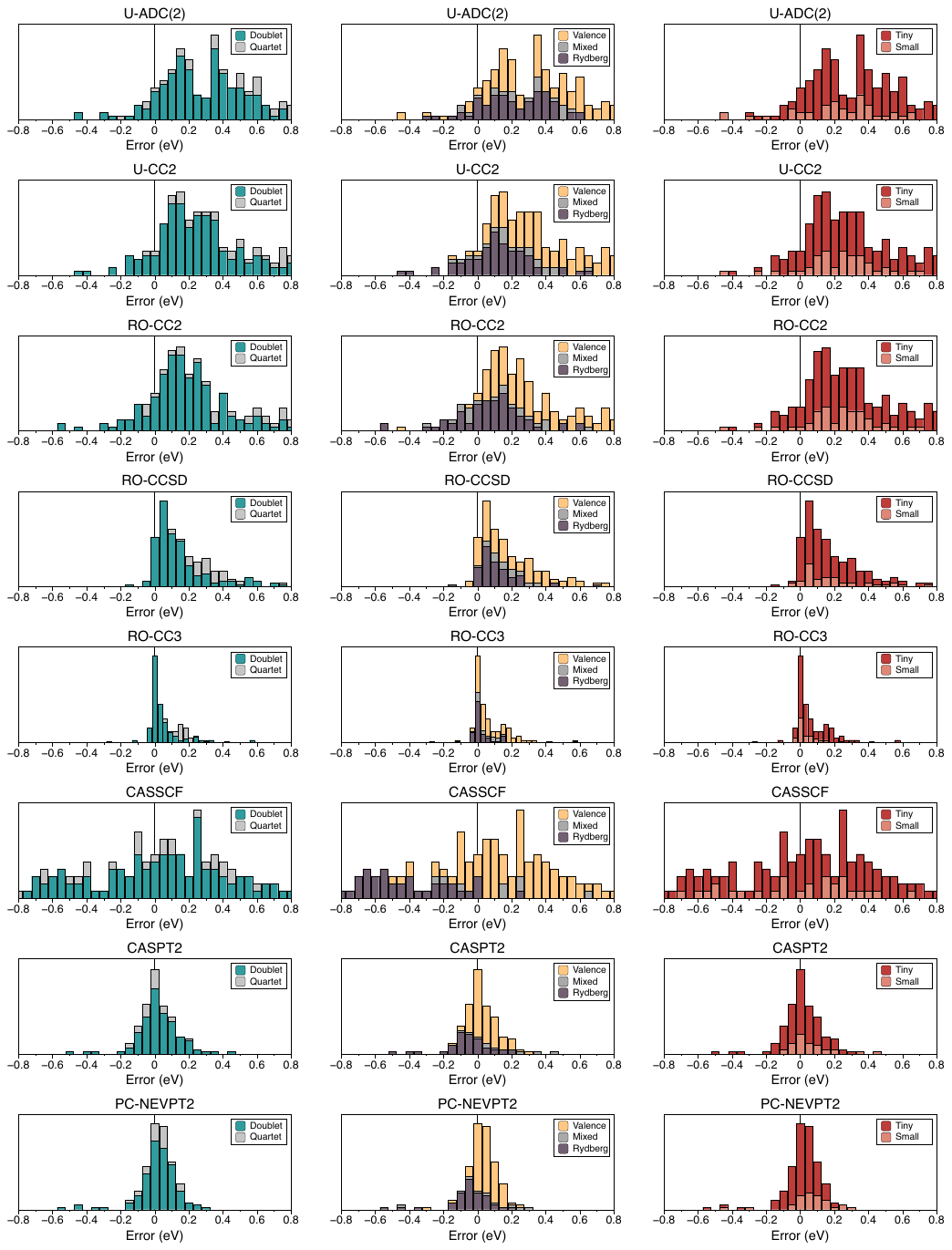}
  \caption{Distribution of the errors on VTEs for various levels of theory  {using stacked histograms and} considering all data from the \textsc{rad} subset, excluding genuine double excitations and unsafe TBEs. From top to bottom:  U-ADC(2),  U-CC2, RO-CC2, RO-CCSD,  ROCC3, CASSCF, CASPT2 (with IPEA), and PC-NEVPT2. From left to right: impact of the spin symmetry, effect of ES nature, and influence of molecular size.}
    \label{Fig-c7}
\end{figure*}

Let us now focus on the radicals in the \textsc{rad} subset, where we have again considered all safe TBEs, excluding the genuine double excitations, to perform our benchmarks. Table \ref{Table-6} presents key statistical data, while Figures \ref{Fig-c7} and S11 to S13 display histograms for selected methods and for all evaluated methods, respectively.

For the CC approaches, both the U and RO formulations were considered. At the CC level, the differences between the results of the two schemes rapidly diminish as the excitation order in CC increases, as expected when approaching the exact solution. Specifically, the mean absolute difference (MAD) between the two open-shell schemes is 0.084, 0.042, and 0.017 eV for CC2, CCSD, and CC3, respectively. For all three CC models, applying RO proves beneficial, though the trends between U-CC and RO-CC results are already similar with CC2, as shown in the second and third rows of Figure \ref{Fig-c7}. This advantage of using RO has also been observed for both ADC(2) and ADC(3) in a recent study. \cite{Sta24} As in the case of closed-shell molecules, notable similarities between the error patterns of U-ADC(2) and U-CC2 are evident in Figure \ref{Fig-c7}, with the CC approach showing a slightly less positively skewed distribution than its ADC counterpart.

When compared to the TBEs, which were established using CCSDTQ, CCSDTQP, or CIPSI, it is immediately apparent that all SR methods yield significantly larger errors for the radicals than for the closed-shell molecules. Additionally, these methods tend to be less accurate for quartet states compared to doublet states, with the former group exhibiting MAEs approximately 50--150\% larger than the latter. This is a logical consequence of the typically stronger multi-configurational character of the electronic states of radicals, particularly in high-spin scenarios.

U-ADC(2) delivers rather poor results, with a significant overestimation (MSE of +0.35 eV) and large dispersion (SDE of {0.32} eV), and MAEs exceeding 0.2 eV for all subgroups of states. As with closed-shell molecules, ADC(3) tends to underestimate transition energies, although U-ADC(3) shows a rather reasonable performance, with a MAE of 0.24 eV, a small MSE, and accurate Rydberg energies. However, it is important to note that state identification is particularly challenging for radicals when using ADC(3), and unambiguous assignments were not achievable in a significant number of cases.

The RO-CC2 MAE for valence transitions, 0.43 eV, is about three times larger than for closed-shell systems. While this value is quite large in our view, comparisons between the tiny and small molecules suggest potential improvements as system size increases. However, it is difficult to confirm this trend, given that the largest systems in the \textsc{rad} subset contain only four non-H atoms. Unsurprisingly, RO-CCSD tends to overshoot transition energies, particularly for valence excited states, though it shows significant improvements compared to RO-CC2 in terms of both MAE and SDE. As expected, climbing to CC3 yields substantial accuracy improvements, but a clear overestimation trend persists. Additionally, the RO-CC3 MAEs of 0.06 eV (doublet) and 0.16 eV (quartet) are much larger than those for the singlet and triplet states (0.02 eV), {and it reaches chemical accuracy in 60$\%$ of the cases only.}

As in the previous section, CASSCF remains relatively untrustworthy, with large MAE (0.37 eV) and SDE (0.35 eV), and overestimated (underestimated) valence (Rydberg) transition energies (see Figure \ref{Fig-c7}). In contrast, all MR approaches that account for dynamic correlation deliver far more satisfying results for the \textsc{rad} states. Applying the IPEA shift improves the CASPT2 values and has again a modest impact on the CASPT3 results. CASPT2, CASPT3, and PC-NEVPT2 all yield MAEs smaller than 0.10 eV for both doublet and quartet states, as well as for both valence and Rydberg transitions, underscoring the effectiveness of these approaches for radicals. Indeed, these three methods provide results comparable to RO-CC3 for doublet states, but they are clearly more effective for quartet states.

\subsection{Comparisons with other benchmark sets}

It may be informative to compare the results obtained herein with those from previous benchmark studies. We focus on singlet transitions in closed-shell compounds, as numerous works have been dedicated to these transitions. Selected values are listed in Table \ref{Table-7}.

\begin{table}[htp]
\caption{\footnotesize Mean signed error (MSE) and mean absolute error (MAE), in eV, reported in selected benchmark studies devoted to singlet ESs. ``Ref.'' indicates the reference used to define the benchmark energies, and ``Method'' denotes the tested theoretical approach.}
\label{Table-7}
\vspace{-0.3 cm}
\footnotesize
\begin{tabular}{lllccr}
\hline
Set				& Ref.		&Method		&MSE$^a$		&MAE$^a$		& Source\\
\hline
Org.~Mol.			&Exp. 0-0	&ADC(2)			&	0.01			&0.08		&\citenum{Win13}\\
				&			&CC2		&	0.04			&0.08		&\citenum{Win13}\\	
				&			&SCS-CC2	&	0.05			&0.06		&\citenum{Win13}\\	
				&			&SOS-CC2	&	0.03			&0.06		&\citenum{Win13}\\	
Org.~Mol.			&Exp. 0-0	&CC2			&	0.04			&0.08		&\citenum{Loo18b}\\		
				&			&CCSD		&	0.21			&0.21		&\citenum{Loo18b}\\			
				&			&CCSDR(3)	&	0.04			&0.05		&\citenum{Loo18b}\\			
				&			&CC3		&	-0.01			&0.02		&\citenum{Loo18b}\\
Thiel	 			& CASPT2	& CC2		& 	0.29	(0.17)	& 0.32 (0.22)	& \citenum{Sch08}\\
				&			&CCSD		&	0.40	(0.36)	& 0.50 (0.37)	&\citenum{Sch08}\\
				&			&CC3		&	0.20 (0.19)	& 0.22 (0.22)	&\citenum{Sch08} \\
				& CC3		& CC2		&	0.13	(0.04)	& 0.17 (0.09)	&\citenum{Sau09} \\
				&			&STEOM$^b$	&	(-0.02)		&(0.07)		&\citenum{Dut18}\\
				&			&CCSD		&	0.30	(0.16)	&0.30 (0.16)	&\citenum{Sau09} \\
				&			&CCSDR(3)	&	0.09 (0.02)	&0.09 (0.02)	&\citenum{Sau09} \\
				& 			&CCSDT-3	&	(0.05)		&(0.05)		&\citenum{Sou14}\\
				&			&ADC(2)		&	-0.03			&0.27		&\citenum{Har14} \\
				&			&ADC(3)		&	-0.20			&0.29		&\citenum{Har14} \\
				&TBE		&ADC(2)		&	0.22			&0.29		&\citenum{Har14} \\
				&			&ADC(3)		&	0.12			&0.24		&\citenum{Har14} \\
				&			&CC3		&	0.23			&0.24		&\citenum{Har14} \\
DNA bases		&CC3		&CC2		&	0.03			&0.06		&\citenum{Kan14b}\\
				&			&CCSD		&	0.11			&0.30		&\citenum{Kan14b}\\	
				&			&CCSDR(3)	&	0.04			&0.09		&\citenum{Kan14b}\\		
Rydberg			&CC3		&STEOM$^b$	&	0.06			&0.07		&\citenum{Dut18}\\
Rydberg			&CCSDT		&CIS(D)		&	-0.23			&0.23		&\citenum{Kan17}\\		
				&			&CC2		&	-0.29			&0.29		&\citenum{Kan17}\\		
				&			&CCSD		&	0.09			&0.09		&\citenum{Kan17}\\	
\hline
\end{tabular}
\vspace{-0.3 cm}
\begin{flushleft}
$^a$ Values in parentheses correspond to selections including only transitions with a CC3 $\%T_1$ greater than 90\% (Refs.~\citenum{Sch08} and \citenum{Sau09}) or greater than 87\% (Refs.~\citenum{Sou14} and \citenum{Dut18});
$^b$ STEOM stands for STEOM-CCSD.
\end{flushleft}
\end{table}

Would comparisons with experimental data yield the same conclusions as \textsc{quest}? Studies relying on experimental values {should, in principle, use} 0-0 energies as references. \cite{Die04b,Win13,Fan14b,Loo18b,Loo19b} It has been shown that theoretical errors on 0-0 energies are primarily driven by inaccuracies in transition energies rather than geometric or vibrational parameters, \cite{Loo19a} so comparisons with the present data are likely valid. In the 2013 study by Winter and collaborators, a MAE of 0.08 eV was reported for CC2, \cite{Win13} and the same MAE was found by us a few years later using a different molecular set. \cite{Loo18b} This 0.08 eV value seems smaller than the one reported in Table \ref{Table-6} for CC2. However, the vast majority of experimentally determined gas-phase 0-0 energies are associated with the first or second singlet ES in medium-sized organic compounds, which biases the 0-0 benchmarks toward low-lying valence transitions. If we analyze \textsc{quest}'s CC2 errors for valence singlet transitions, we obtain MAEs of 0.10 (0.08) eV for medium (large) compounds, values that are consistent with 0-0 benchmarks. The same holds for CCSD: the MAE of 0.21 eV reported in Ref.~\citenum{Loo18b} might appear excessive, but we obtain similar MAEs of 0.24 (0.25 eV) for valence singlet ESs in medium (large) compounds of  \textsc{quest}.

It is also natural to consider benchmarks performed with Thiel's set, which, as a reminder, focuses on valence transitions only. \cite{Sch08,Sau09} Interestingly, in these works, the authors emphasized that the most significant aspect of their approach was comparing methods on a perfectly equal footing. For this reason, they did not use their TBEs for benchmarking, as some values were sourced from the literature (\emph{vide supra}). Instead, they preferred comparing methods with CASPT2/TZVP (CC3/TZVP) results for singlet (triplet) transitions. \cite{Sch08} As can be seen in Table \ref{Table-7}, this approach leads to large MSE and MAE values for all three CC schemes for the singlet ESs. However, there is a significant decrease in errors for both CC2 and CCSD when considering only transitions where $\%T_1 > 90\%$. In a subsequent work focused on CCSDR(3), \cite{Sau09} the same group used CC3 as a reference, yielding statistical values for the various CC approaches that are relatively close to the present estimates. However, the reported CCSD error (MSE and MAE of 0.30 eV) still appears significantly larger than ours when accounting for all states.

The 2014 analysis in Ref.~\citenum{Har14}, based on Thiel's data, is also very informative. In that work, the authors compared ADC(2) and ADC(3) transition energies with both Thiel's TBEs and CC3 values. Using the former as reference, they found nearly equivalent MAEs for ADC(3) and CC3, with a smaller MSE for the ADC(3) approach. This led them to conclude that it was not possible to definitively determine which of the two methods was more accurate. However, choosing the CC3 values as references logically leads to a conclusion more in line with the one we report here: ADC(3) tends to significantly undershoot the transition energies. {It only became clear more recently that CC3 VTEs are more reliable than their CASPT2 counterparts for singlet states}.

Kannar, Tajti, and Szalay provided a benchmark on small molecules aimed at comparing basis set effects for both valence and Rydberg transitions, \cite{Kan17} and we report in Table \ref{Table-7} results for Rydberg transitions with the AVTZ basis. The sign of the MSEs is consistent with what we found here for Rydberg transitions, showing significant underestimations with both CIS(D) and CC2, and a slight overestimation with CCSD. However, the magnitudes of both MSE and MAE are larger in Ref.~\citenum{Kan17} than here for the two former approaches. Specifically, for CIS(D), the reported MSE of $-0.23$ eV seems inconsistent with the $-0.05$ eV observed in Table \ref{Table-6}. If we limit our analysis of \textsc{quest} results to the tiny compounds, we obtain a MSE of $-0.07$ eV for the Rydberg singlets with CIS(D). We therefore hypothesize that the largest deviations in Kannar \textit{et al.}'s study are due to the consideration of many high-lying Rydberg ESs.

In summary, as expected, reference and test sets play a crucial role in assessing methods, and one must exercise caution when extrapolating the results of specific benchmark studies. In this context, the tools provided in the \textsc{GitHub} repository enable interested readers to obtain error estimates for any subset of their choice.

\subsection{\textsc{diet sets}: an optimized path to accuracy}

When developing new theoretical approaches, especially those with non-optimized or computationally intensive implementations, benchmarking against large-scale datasets can be time-consuming. To address this, one can build `diet' subsets: compact yet statistically representative selections designed to mirror the key properties of the full dataset. These subsets are ideal for quickly evaluating new methods or for training machine learning models when computational resources are constrained.

An illustrative precedent is the diet GMTKN55 set proposed by Gould, \cite{Gou18} distilled from the comprehensive GMTKN55 suite \cite{Goe17} (see also Ref.~\citenum{Gou25}). In the same spirit, our \textsc{GitHub} repository includes a Python-based script to generate diet subsets of the \textsc{quest} database. This tool can likely be especially useful during early-stage development or exploratory testing of computational methods.

The repository supports generation of optimized subsets of varying sizes, e.g., 50, 100, or 200 excitations, using a genetic algorithm that selects transitions in order to closely match the full dataset's statistical distribution. The selection is guided by a multi-objective scoring function that balances the MSE, MAE, and RMSE across multiple SR wave function methods. By default, multireference methods such as CASSCF, CASPT2, and NEVPT2 are excluded from the scoring, but the script can be easily adapted to include or prioritize different approaches.

Key features of the toolset include:
(i) generation of statistically representative subsets that reflect the full dataset's composition in terms of spin states (singlet \textit{vs} triplet), nature of the transitions (valence \textit{vs} Rydberg), excitation types (e.g., $\npi$, $\ppi$, etc.), molecular size, etc.;
(ii) support for flexible, user-defined filtering options (e.g., only singlets, only valence transitions, exclusion of genuine double excitations); and
(iii) preservation of full excitation metadata in the output JSON files, enabling seamless post-processing or downstream analysis.

As an example, we provide on the \textsc{GitHub} repository and in  Section S4 of the SI a 50-excitation subset spanning 20 molecules, derived from the filtered \textsc{main} dataset (excluding unsafe TBEs and genuine double excitations). As can be seen, it provides accurate reproduction of the main statistical trends. We also checked that the same holds for the singlet and triplet subgroups of the methods including in the training. We also underline that this subset is optimized to reflect the statistical structure of the full set, yet it is not unique due to the stochastic nature of the genetic algorithm. Multiple near-optimal subsets may exist, offering flexibility in applications where diversity is desirable. 

\section{Conclusions and Perspectives}

Benchmark datasets and their associated reference values have long served as a cornerstone for advancing electronic structure theory, playing a central role in the validation, calibration, and continuous refinement of computational methods. \cite{Pop89,Cur91,Cur97,Cur98,Taj04,Bom06,Jur06,Zha06i,Cur07b,Har08,Goe10,Goe11,Goe11b,Rez11,Van15,Mot17,Mar17c,Goe17,Stu20,Wil20}  In the specific context of ES quantum chemistry,  \cite{Fur02,Die04,Die04b,Sch08,Sil08,Goe10a,Sil10,Sil10b,Sil10c,Sen11,Sen11b,Sen11c,Jac12c,Lea12,Win13,Jac15a,Jac15b,Sch17,Cas19,Cas20,Koz20,Son25} such datasets are indispensable for providing a systematic and objective assessment of the performance of both wave function and density-based methods across a chemically diverse set of transitions. The accuracy and reliability of theoretical predictions critically depend on well-defined benchmark protocols and high-quality reference values that allow one to disentangle methodological strengths and limitations.

In this work, we have pursued and expanded upon this philosophy through the development of the \textsc{quest} dataset, a curated collection of high-level VTEs for a broad and diverse set of molecular systems. The dataset incorporates TBEs derived from systematically converged wave function methods, including SCI and high-order CC calculations, typically performed in triple-$\zeta$ basis sets and corrected when feasible with larger basis sets using additive schemes. The use of a consistent computational protocol, combined with the deliberate exclusion of experimental values as references, ensures internal consistency and eliminates potential biases linked to experimental uncertainties or solvent/environmental effects.

Importantly, though we did not detail this aspect herein, the \textsc{quest} initiative goes beyond excitation energies. It provides chemically accurate 0-0 transition energies \cite{Loo18b,Loo19a,Loo19b} and charged (i.e., electron detachment) excitations. \cite{Mar24,Mar25} In recent years, it has been extended to include other ES properties such as transition dipole moments, oscillator strengths, and two-photon crosssections. \cite{Chr21,Sar21,Nai24,Sir25} These additions are essential for the benchmarking of methods aimed at simulating spectroscopic observables and for providing a more complete theoretical picture of ES behavior. The dataset also now encompasses more challenging electronic excitations, such as those in radicals \cite{Loo20d} and double-excitation-dominated states, \cite{Loo19c,Kos24} where standard single-reference methods often fail. The data of \textsc{quest} are accessible through a dedicated \textsc{GitHub} repository that can be found at the following URL: \url{https://github.com/pfloos/QUESTDB}. 

The adoption of the \textsc{quest} dataset by many research groups highlights its emergence as a useful tool for ES electronic structure theory. \cite{Gin19,Cas19,Oti20,Hai20b,Hai21,Gou21,Gro21,Mes21,Cas21b,Mes21b,Lia22,Mes22,Gou22,Kin22,Van22,Fur23,Hen23,Kin23,Gro23,Ent23,Ris23,Zie23,Cas24,Cos24,Pfa24,Sza24,Sta24,Dat24,Gou24,Bin25,For25,Mil25,Tra25,Yu25,Son25,Val25}  
It not only helps identifying which methods are reliable under which conditions, but also reveals the boundaries where established approaches begin to break down, helping guiding both the selection of appropriate computational tools and the design of new methodological innovations. By expanding the chemical space and including electronically and spectroscopically demanding cases, the present extension of \textsc{quest} will hopefully be considered valuable by the community.

Through ongoing efforts to enrich its content with larger molecules, additional electronic properties, and more diverse chemical environments, it is our hope that \textsc{quest} could continue to support the critical task of benchmarking. It is also possible that this dataset will catalyze the development of predictive models capable of generating new insights into ES behavior thanks to AI-driven developments. \cite{Pfa24}

\section*{Acknowledgements}
The authors are deeply indebted to all PhD students, post-doctoral associates, and collaborators that directly contributed to  \textsc{quest} since 2018.
{DJ thanks B\'eatrice Lejeune ($@$bea$\_$quarelle on Instagram) for the TOC.}
PFL thanks the European Research Council (ERC) under the European Union's Horizon 2020 research and innovation programme (Grant agreement No.~863481) for funding.
FL acknowledges financial support from ICSC-Centro Nazionale di Ricerca in High Performance Computing, Big Data, and Quantum Computing, funded by the European Union 
- Next Generation EU - PNRR, Missione 4 Componente 2 Investimento 1.4.
This work used the HPC resources from CALMIP (Toulouse) under allocations 2024-18005 and 2025-18005, as well as resources from the GLiCID Computing 
Facility (\emph{Ligerien Group for Intensive Distributed Computing}, https://doi.org/10.60487/glicid, Pays de la Loire, France). 

\section*{Supporting Information Available}
Extra analyses, plots and data (.pdf). Full list of raw data (.xls). Cartesian coordinates for all compounds (.pdf).

\bibliography{biblio-new}

\end{document}